\documentclass[structabstract]{aa}    
\usepackage{amssymb,amsmath,mathabx}
\usepackage{graphicx}  
\usepackage{txfonts}
\usepackage{natbib} 
\usepackage{color}   
\usepackage{threeparttable}  
\usepackage{booktabs}
\usepackage{wasysym}
\bibpunct{(}{)}{;}{a}{}{,} 

\def\ProDiMo{{\sc ProDiMo}}
\def\phyllos{{phyllosilicate}}
\def\Phyllos{{Phyllosilicate}}
\def\H2O{H$_2$O}
\def\pd{protoplanetary disk}
\def\sn{Solar Nebula}
\def\physi{physisorbed\ }
\def\Ncorelay{$N_{\rm core,layer}$}  
 \def\LH{Langmuir-Hinshelwood}
 \def\ER{Eley-Rideal}

\def\rra{$\rightarrow$}

\def\H2{H$_{\rm 2}$}

\def\UV{\rm UV}
\def\nd{n_{\rm d}}
\def\Td{T_{\rm d}}

\def\Tg{T_{\rm g}}

\def\Ns{N_{\rm surf}}

\def\therdes{^{\rm des, th}_i}

\def\EbHp{$E^{\rm b}_{\rm H\varhash}$}

\def\EbHc{$E^{\rm b}_{*{\rm H\varhash}}$}

\def\EbHPAH{$E^{\rm b}_{\rm HPAH}$}

\def\EdHHPAH{$E^{\rm diss}_{\rm H_2,PAH}$}

\def\EdHHchem{$E^{\rm diss}_{\rm H_2,*}$}

\def\act{^{\rm act}_i}

\def\actPAHH{^{\rm act}_{\rm PAH,H}}

\def\actPAHpH{^{\rm act}_{\rm PAH^+,H}}

\def\actHpHp{^{\rm act}_{\rm H\varhash,H\varhash}}
\def\actHcHc{^{\rm act}_{\rm *H\varhash,*H\varhash}}

\def\actHpHc{^{\rm act}_{*{\rm H\varhash,H\varhash}}}

\def\Eactc{E^{\rm act}_{\rm *}}

\def\gcH{^{\rm gc}_{\rm H}}

\def\pcH{^{\rm pc}}

\def\freq{\nu_{0,i}}

\def\nHtot{n_{\langle {\rm H} \rangle}}

\def\cm3{\ \rm cm^{-3}}  
  
\def\surfchem{_{\mathrm{surf,chem}}}

\def\1M{{\sc MC-Analytic}}
\def\2M{{\sc MC-Phys}}
\def\3M{{\sc MC-Phys-Chem-PAH}}
\def\4M{{\sc Disk-analytic}}
\def\5M{{\sc Disk-Phys}}
\def\6M{{\sc Disk-Phys-Chem-PAH}}

\def\a2{a^2}

\def\UV{$h\nu$} 

\begin{document}

\title{Warm dust surface chemistry in protoplanetary disks}
\subtitle{Formation of \phyllos s}
\author{W. F. Thi\inst{1}, S. Hocuk\inst{1,6}, I. Kamp\inst{2}, P. Woitke\inst{3,7}, Ch. Rab\inst{2}, S. Cazaux\inst{4}, P. Caselli\inst{1}, M. D'Angelo\inst{5,2}}
\institute{
Max Planck Institute for Extraterrestrial Physics, Giessenbachstrasse, 85741 Garching, Germany 
\and 
Kapteyn Astronomical Institute, University of Groningen, Postbus 800,
             NL-9700 AV Groningen, The Netherlands
\and             
SUPA, School of Physics \& Astronomy, University of St. Andrews, North Haugh, St. Andrews, KY16 9SS, UK
\and
Faculty of Aerospace Engineering, Delft University of Technology, Delft, The Netherlands
\and
Zernike Institute for Advanced Materials, University of Groningen, P. O. Box 221, 9700 AE Groningen, The Netherlands
\and
CentERdata, Tilburg University, P. O. Box 90153, 5000 LE, Tilburg, The Netherlands
\and
Centre for Exoplanet Science, University of St Andrews, St Andrews, UK
}

  \abstract
  {The origin of the reservoirs of water on Earth is debated. The Earth's crust may contain at least three times more water than the oceans. This crust water is found in the form of \phyllos s, whose origin probably differs from that of the oceans.}
  {We test the possibility to form \phyllos s in \pd s, which can be the building blocks of terrestrial planets.}
  {We developed an exploratory rate-based warm surface chemistry model where water from the gas-phase can chemisorb on dust grain surfaces and subsequently diffuse into the silicate cores. We apply the \phyllos\ formation model to a zero-dimensional chemical model and to a 2D \pd\ model (\ProDiMo). The disk model includes in addition to the cold and warm surface chemistry continuum and line radiative transfer, photoprocesses (photodissociation, photoionization, and photodesorption), gas-phase cold and warm chemistry including three-body reactions, and detailed thermal balance.} 
  {Despite the high energy barrier for water chemisorption on silicate grain surfaces and for diffusion into the core, the chemisorption sites at the surfaces can be occupied by a hydroxyl bond (-OH) at all gas and dust temperatures from 80 to 700~K for a gas density of 2$\times$10$^{4}$ $\cm3$. The chemisorption sites in the silicate cores are occupied at temperatures between 250 and 700~K. At higher temperatures thermal desorption of chemisorbed water occurs. The occupation efficiency is only limited by the maximum water uptake of the silicate. The timescales for complete hydration are at most 10$^5$ years for 1 mm radius grains at a gas density of 10$^{8}$ cm$^{-3}$.}   
  {Phyllosilicates can be formed on dust grains at the dust coagulation stage in \pd s within 1 Myr. It is however not clear whether the amount of \phyllos\ formed by warm surface chemistry is sufficient compared to that found in Solar System objects.}

\authorrunning{W. F. Thi}
\titlerunning{Phyllosilicates  in \pd s}
\maketitle

%

\section{Introduction}\label{introduction}

The origin of the water reservoirs on Earth is a long-standing debate \citep{vanDishoeck2014prpl.conf..835V,Ceccarelli2014prpl.conf..859C,Caselli2012A&ARv..20...56C,Drake2015_MAPS:MAPS960,Cleeves2014Sci...345.1590C}. What were the sources of water in terrestrial planets? Was water contained in the Earth's building blocks already at its formation or was the water delivered later during the late heavy bombardment phase? 

The Earth's bulk composition and that of meteorites known as enstatite chondrites suggest a ''dry'' proto-Earth scenario: the initial building blocks of the Earth are made of dry rocks, composed only of non-hydrated silicates and carbonaceous materials. Most of the water is brought afterward by comets and asteroids during the phase of heavy bombardments \citep{Nesvorny2017AJ....153..103N}. In addition to the water in the oceans, large reservoirs of water trapped in silicates in the Earth's mantle have been found with up to three times the amount of water in Earth's oceans \citep{Schmandt2014,Marty2012E&PSL.313...56M}. The origin of the water in the Earth's mantle may differ from that of the oceans and the amount of mixing between these reservoirs is not clear. This scenario is challenged by the Rosetta mission. Rosetta measurements found that the Jupiter family comet 67P/Churyumov-Gerasimenko has a much higher HDO/H$_2$O ratio that in the Earth's oceans \citep{Altwegg1261952}.

In the second model, the so-called ''wet'' scenario, most of the water on Earth comes from the release of water vapour trapped inside the planetesimals like carbonaceous chondrites as hydrated silicate upon impact or during volcanism \citep{vanDishoeck2014prpl.conf..835V,Drake2005M&PS...40..519D}. Water is already present in the initial phase of Earth's formation.

Water-rich planetesimals located at 2--3 au are perturbed by the giant planets and collide with the young Earth \citep{Morbidelli2000M&PS...35.1309M,Gomes2005Natur.435..466G,Raymond2004Icar..168....1R,Raymond2005ApJ...632..670R}. Such planetesimals contain water in form of hydrated silicates (\phyllos s). A few main belt asteroids are ice-covered planetesimals, which may have brought water on Earth \citep{Jewitt2012AJ....143...66J}. C-type main belt asteroids, whose spectra resemble those of  carbonaceous chondrite meteorites are located throughout the main belt with the peak distribution located beyond the Solar Nebula snow line \citep{Rivkin2002aste.book..235R}. 

\cite{Alexander2018SSRv..214...36A} reviewed the water reservoir in the Solar System small bodies and \cite{DELOULE19954695} studied the amount of interstellar water in meteorites. Chondrites are known to have kept the record of Solar Nebular and asteroidal histories. Hydrated carbonaceous chondrites are meteorites where water is found in the form of \phyllos s \citep{Beck2010,Beck2014}. It is generally believed that the direct hydration of anhydrous silicates by water vapour in the \sn\ was too inefficient compared to its lifetime \citep{Fegley2000}, although it is thermodynamically possible \citep{ZOLOTOV2012713}. Therefore, it is generally accepted that the original anhydrous silicate in the parent bodies interacted with water-rich fluids leading to the hydrolysis of the silicate \citep{Brearley2006,Bischoff1998M&PS...33.1113B}. However, studies of the D/H ratios in meteorites suggest that water in meteorites can be formed out of thermodynamic equilibrium \citep{DELOULE19954695}.

Original analyses of the Moon's samples implied a dry Moon formation but recent studies suggest that a wet formation scenario is also possible \citep{Saal2013_17,Hui2013NatGe...6..177H,Anand2014RSPTA.37230254A,Chen2015E&PSL.427...37C}. Near-infrared reflectance spectra of the Moon pyroclastic deposits support  the presence of Earth-like water content in the lunar interior \citep{Milliken2017}.

The water in the Earth mantle may have been incorporated already in the dust grains in the form of \phyllos s. This wet scenario is supported by the presence of a lower D/H ratio in lavas than in Earth's oceans \citep{Hallis2015}. The D/H ratio found in lavas may be more representative of the primordial D/H  than the value found in the Earth ocean. The lavas D/H ratio is similar to that of carbonaceous chondrites, which may have acquired their water by absorption at the solar nebula stage.

Hydration at silicate grain surfaces is an alternative mechanism that has been shown to occur experimentally \citep{Rietmeijer2004,Yamamoto2016LPI....47.1733Y,LEGUILLOU2015179}. As an activated process, the hydration timescale depends exponentially on the activation energy and temperature. \citet{Yamamoto2016LPI....47.1733Y} noticed that hydration of amorphous silicates is enhanced compared to the hydration of crystalline silicates. A low activation energy would permit hydration of the small grains in the early life of \pd s. Those hydrated grains will subsequently coagulate into planetesimals \citep{Ciesla2005}. Alternatively, warm gas in the inner region of \pd s or gas heated by shocks has enough energy to overcome the barrier \citep{Ciesla2002M&PSA..37Q..34C,Furukawa2011_6461}. The dust grain hydration models have only explored formation via the \ER\ process, i.e.\ the direct formation of the \phyllos\ bonds upon impact of a gas-phase water with an anhydrous silicate surface, the so-called Simple Collision Theory. The Simple Collision Theory does not account for Langmuir-Hinselwood type reactions between chemisorbed species. Based on the Simple Collision Theory and high values for the activation energy, direct hydration has been considered too inefficient \citep{Fegley2000}.

In the third scenario, gas and dust material from the \pd\ can be incorporated to the young planet Earth atmosphere after the core formation; this is the so-called nebular ingassing model \citep{Sharp2017,Genda200842}. \citet{Wu2018JGRE..123} estimate that a small amount of ($>$0--0.5 oceans) of nebular hydrogen, is generally required, supplementing the chondritic contributions. 

We explore in this paper the ''wet'' formation scenario, i.e.\ the possible hydration process of the dust grains in \pd s using a surface chemistry model adapted to warm dust grains and recent estimates of the activation barriers. Especially we estimate the timescale of \phyllos s formation and the amount of water trapped in silicates.
protoplanetary
\Phyllos s (hydrated silicates) are silicates with a chemically bonded hydroxyl (-OH) function. \Phyllos s can also be considered as OH attached to the silicate at chemisorption sites. In addition to the \ER\ process, i.e.\ the direct formation of chemisorbed OH silicate bonds upon water vapour molecules encounters with silicate grains, we also considered hydration of silicate grain surfaces via the \LH\ process.  In the \LH\ process, the weakly adsorbed water molecules (physisorbed water) scan the silicate surface to search for an empty chemisorption site and attempt to overcome either thermally or via tunnelling the activation barrier. Dust grains at temperatures below 100~K are covered by up to a few hundreds layers of water ice. The amount of water molecules is large with one layer of water ice corresponding for a 0.1~$\mu$m radius grain to $\sim$1.9$\times$10$^6$ molecules.

Water ice is dynamic and mobile \citep{Ghesquiere2015C5CP00558B} and therefore, the process of water diffusion described above is akin to the so-called aqueous alteration \citep{Zolensky2008} although water has an amorphous solid structure instead of being liquid. Even for dust grains that are above the sublimation temperature of water ice,  water molecules would have ample time to scan the surface before they desorb because the experimental and theoretical value of the activation energy to form a chemisorption bond is around 3000~K \citep{Wegner1983,Stimpfl2006,Muralidharan2008,Leeuw2010,King2010,Prigiobbe2013,Asaduzzaman2014E&PSL.408..355A}. This is below the desorption energy of \physi water \citep[$\sim$~5700~K,][]{Fraser2001MNRAS.327.1165F}.  

\cite{DAngelo2018arXiv180806183D} used a Monte Carlo numerical code to study in details the adsorption of water on silicate between 300 and 800~K. In their Monte-Carlo simulation the [100] forsterite crystal lattice is simulated by a grid composed of 20 $\times$ 20 cells. Each unit cell has four possible binding sites with binding energies ranging between 8000 and 20000 K corresponding to Mg cations, three of which are closer to the surface and easily accessible to water molecules. The highest binding sites represent 45\% of the total number of sites. Also, the formation of water clusters on surfaces have been investigated, and showed that clusters increase the water coverage of the dust. By considering small dust particles of 0.1 microns, an important amount of water of up to 10 earth oceans can be available in the early earth. 

\citet{Woitke2018A&A...614A...1W} performed a chemical equilibrium calculation and found that for inner midplane protoplanetary disk conditions phyllosilicates are the most thermodynamically stable species. 

In this study we focused on the global hydration kinetics of silicate grains in protoplanetary disk conditions. Another novelty in our study is that we took the diffusion into the silicate core into account.

The paper is organised as follows: we first introduced the chemical model in Sec.~\ref{chemical_model}; the code and the models are presented in Sec.~\ref{modelling}; the results are shown and discussed in Sec.~\ref{results} and \ref{discussion}; the conclusions and perspectives are given in Sec.~\ref{conclusions}.

\section{Silicate hydration and dehydration model}~\label{chemical_model} 
\begin{table}
\begin{center}
\caption{Special species. \label{tab_species}}
\vspace*{-0.5mm}          
\resizebox{88mm}{!}{\begin{tabular}{lcc}     
\toprule
Name & Symbol & Remark\\ 
\noalign{\smallskip}     
\hline
\noalign{\smallskip}  
free silicate surface chemisorption site & * & pseudo element/species\\
free silicate core chemisorption site & ! & pseudo element/species\\
H gas    & H & \\
\H2 gas  & H2 & \\
silicate surface chemisorbed H & *H\# & \\
\H2O gas & H2O & \\
physisorbed (ice) \H2O &H2O\# & \\
silicate surface chemisorbed \H2O & *H2O\# & \\
silicate core chemisorbed \H2O & !H2O\# & \\
\noalign{\smallskip}
\bottomrule
\end{tabular}}
\vspace*{-2mm} 
\tablefoot{[*]+[*H\#]+[*H2O\#] and [!]+[!H2O\#] are conserved values. Pseudo-species do not have mass.}
\end{center}
\end{table}

Chemisorption concerns two species in our model atomic hydrogen and water. Chemisorbed H atoms (on the silicate surface *H\#) can recombine with another H atom to form \H2. Details on the \H2 formation in our model is explained in a previous paper (Thi et al., in submitted). The formation occurs on the grain surfaces by the \LH\ and \ER\ processes involving \physi and chemisorbed H atoms. \H2 can also form via hydrogenated PAHs (HPAH) and PAH cations (HPAH$^+$). Both the H atoms and water molecules are assumed to compete for the available number of surface chemisorption sites (at the silicate surface and in the core). More sophisticated models take the possibility of a species with a high binding energy to swap pace with less strongly bounded species into account \citep{Cuppen2007ApJ...668..294C}. We assumed in this exploratory study that only hydrogen atoms and water molecules can diffuse from the silicate surface inside the silicate core. The adopted surface density of chemisorption sites $N_{\rm surf}$ is 1.5$\times$10$^{15}$ cm$^{-2}$. We modelled only five and 500 layers inside the silicate core for the cloud model and \pd\ model respectively in addition to the silicate surface layer, making a total of 6 and 501 layers of chemisorption sites for hydrogen atoms and water to bind to. The choice for 500 silicate core layers will be discussed in a subsequent section.

Water can form in the gas-phase by ion-neutral reactions at low temperatures and via neutral-neutral reactions at high temperatures \citep{Thi2005A&A...438..557T,Thi2010MNRAS.407..232T,vanDishoeck2014prpl.conf..835V}. At temperatures below the sublimation temperature of the water ice, water is formed on the surface by encounters of \physi species starting from the hydrogenation of O into OH. Hydrogenation of atomic surface atoms is efficient below $\sim$~15~K.

We restricted ourselves to this surface path although alternative routes via O$_2$H\# and H$_2$O$_2$\# for example have been proposed \citep{Taquet2013A&A...550A.127T,Cuppen2017SSRv..tmp....2C,Lamberts2014A&A...570A..57L,Cazaux2010A&A...522A..74C}. At high temperatures water is first formed in the gas phase before being incorporated in the grains. Another important route is the reaction of OH\# with \H2\# \citep{Penteado2017ApJ...844...71P}.

We updated the gas-phase reaction rate coefficients between OH and \H2 with the values from \citet{Meisner2016JChPh}, who took tunnelling effects into account in their computations. Warm gas-phase chemistry of water is discussed in the context of \pd s in \citet{Thi2010MNRAS.407..232T,Kamp2013A&A...559A..24K,Antonellini2015A&A...582A.105A,Antonellini2016A&A...585A..61A,Woitke2009A&A...501L...5W,vanDishoeck2014prpl.conf..835V,Bethell2009Sci...326.1675B,Du2014ApJ...792....2D}.

Water ice molecules located at the interface between the icy mantle and the silicate surface can overcome an activation barrier to chemisorb to the silicate surface. Likewise, if the grain is too hot to host an icy mantle, water molecules can impinge directly on the silicate surface, scan it, find a suitable free chemisorption site, and attempt to overcome the barrier. If the dust is warm enough, the chemisorbed water can diffuse inside the silicate core.

We used the theoretical activation energy from \cite{Asaduzzaman2014E&PSL.408..355A} for the formation of a chemisorption bond between water (previously physisorbed or in the gas-phase) and the silicate (3010~K), consistent with experimental values in the order of 2500~K \citep{Yamamoto2016LPI....47.1733Y}. We chose a binding energy of 32710~K for water chemisorption, consistent with dehydration experiments of \phyllos\ \citep{Sawai2013}. The water diffusivity in glasses has been measured to have an activation energy of 13470~K for low water concentration in amorphous silicon dioxide \citep{Okumura2004}. \citet{Kostinski2012IEDL...33..863K} measured a value of 12770~K for the water vapour penetration barrier in the silicate core. The diffusion over chemisorption energy is 0.39. The desorption and diffusion are calculated according to standard cold surface chemistry methods \citep{Hasegawa1993MNRAS.261...83H}. 
The activation barrier for the formation of a chemisorption bond is much lower than the barrier for water diffusion from a chemisorbed water at the silicate surface to another chemisorption site in the silicate core. Therefore, we expect the formation of a layer of chemisorbed water on the silicate surface at much lower dust temperature than the diffusion of chemisorbed silicate surface water into the silicate core. This is called a precursor-mediated chemisorption, the physisorption site being a precursor state. The desorption energy is simply the binding energy in the chemisorbed state.

We adopted the same value for diffusion of water on the silicate surface and for diffusion from the silicate surface towards the core centre. 
The desorption rate pre-factor frequency follows the formula for a rectangular barrier of width $a_{\rm c}$ and height $E\therdes$
\begin{equation}
\freq=\sqrt{\frac{2 \Ns E\therdes}{\pi^2 m_i}}.
\end{equation}
We derived a frequency $\freq$ of (1--10) $\times$ 10$^{12}$ Hz assuming a surface site density is $\Ns$ = 1.5 $\times$ 10$^{15}$ sites cm$^{-2}$. This formula may underestimate the actual values up to 10$^{16}$ Hz \citep{Rettner1996}.

The water molecules and H atoms compete for the limited silicate surface and silicate core chemisorption sites. A chemisorbed H-atom can recombine with another H-atom to form \H2, while a hydroxyl bond is assumed to be chemically inert once formed. The chemisorbed species and some special reactions included in our model are summarised in Table~\ref{tab_species} and \ref{tab_reactions}. 

\section{Modelling}\label{modelling}

\begin{table}
\begin{center}
\caption{Main grain reactions involved in the formation and destruction of chemisorbed water.\label{tab_reactions}}
\vspace*{-3mm}          
\begin{tabular}{p{0.2cm}p{0.5cm}p{0.2cm}p{0.5cm}p{0.2cm}p{0.5cm}p{0.2cm}p{0.5cm}p{1.5cm}}     
\toprule
\noalign{\smallskip}        
\multicolumn{8}{c}{Reaction} & \multicolumn{1}{c}{Energy}\\
\hline
\noalign{\smallskip}
1 &  \H2O  & + & *   & \rra & *\H2O\#    & & &   $E\act$~   =~3010~K\\
2 &  \H2O\#  & + & *   & \rra & *\H2O\#    & & &  $E\act$~   =~3010~K\\
3 &  *\H2O\#   & &  & \rra & H2O    & * & &  $E\therdes$~=~32710~K \\
4 &  *\H2O\#  & + & !   & \rra & !\H2O\#    & + & * &   $E\act$~   =~13470~K \\
5 &  !\H2O\#  & &   & \rra & H2O    & + & !  & $E\therdes$~   =~32710~K \\
6 &  \H2O\#   & &   & \rra & \H2O   &   &   & $E\therdes$~   =~5700~K \\
\noalign{\smallskip}
\bottomrule  
\end{tabular}
\tablefoot{$E\act$ means activation energy. $E\therdes$ means desorption energy}
\end{center}
\end{table}

\subsection{The \ProDiMo\ code}\label{ProDiMo_code}

\ProDiMo\ is a hydrostatic 2D code designed to model the physics and chemistry of \pd s \citep{Woitke2016A&A...586A.103W,Woitke2009A&A...501..383W,Kamp2010A&A...510A..18K,Thi2011MNRAS.412..711T,Aresu2011A&A...526A.163A}. It includes detailed continuum and line radiative transfer, heating and cooling balance, and gas and surface chemistry. The gas and surface chemistry is computed by solving rate equations both using a time-dependent or a steady-state solver.
The gas densities in protoplanetary disks are higher than 10$^{4}$ cm$^{-3}$, which makes the use of the rate-equation method appropriate.

\subsection{Chemical network}~\label{chemical_network}

We considered a simplified chemistry in the gas-phase, on the surface of the icy grain mantle, the ice mantle, the surface of the silicate, and the silicate core. The adsorption, diffusion, and desorption rates depend on the actual adsorption sites. However, the model does not account for detailed location of each solid species. The gas-phase and simple ice species are the same as those in the small network described in \cite{Kamp2017A&A...607A..41K} based on the UMIST2012 chemical network \citep{McElroy2013A&A...550A..36M}. The additional species are adsorbed hydrogen atoms and OH ice.

The chemical network is run using the method of "pseudo" chemical species used for the surface chemisorption sites (first "pseudo" element *) and for the silicate core sites (second "pseudo" element !). The ''elemental abundance'' of pseudo elements are set by the total quantity of the chemisorption sites (at the silicate surface and in the silicate core). Assuming a surface site number density $N_{\rm surf}$ of 1.5$\times$10$^{15}$ cm$^{-2}$, the number of silicate surface sites is $N_{\rm surf} \pi \a2$ per grain where $a$  is the grain radius. The number of silicate core chemisorption sites is \Ncorelay $N_{\rm surf} \pi \a2$ per grain, where \Ncorelay\ is a free parameter to account for the amount of ''layers' of ''water'' that a silicate grain can have. Atomic hydrogen is allowed to be physisorbed (H\#) or chemisorbed on the silicate surface (*H\#) while water can be physisorbed (\H2O\#), chemisorbed at the silicate surface (*\H2O\#) or in the silicate core (!\H2O\#). The pseudo elements/species method permits the code to track how a species is bound (physisorbed, chemisorbed at the silicate surface, or chemisorbed in the silicate core) statistically. The method naturally accounts for the limited number of chemisorption sites.

Surface rates account for the competition between diffusion and reaction but not with the desorption processes (thermal, photoabsorption induced and cosmic ray induced desorption), whose effects are explicitly accounted for in the rate equation scheme. Both atomic hydrogen and water can be chemisorbed on the silicate core while only water can chemisorb in the silicate core. There is no reaction-induced desorption. Tunnelling is accounted for diffusions and reactions for all the species assuming a rectangular shape for the barrier and using the Bell's formula \citep{Bell1980}.

The amount of water trapped as \phyllos\ is not kinetically limited but is rather limited by the capacity of the silicate to accommodate large amounts of water. We can estimate the uptake of water in the silicate core as function of the number of layers in the core.  The number density per dust layer is  $n\surfchem=4\pi \Ns \a2 \nd$, which is the number of chemisorption sites per layer. 

When all the chemisorption sites are occupied by water, the total mass of chemisorbed water is
$(1+N_{\rm core,layer})4\pi \Ns \a2 \nd m_{\rm H_2O}$ (g cm$^{-3}$), where the total number density of dust grains is
\begin{equation}\label{eq_ndust}
n_{\mathrm{d}} = \frac{1.386\ \mathrm{amu}\ \nHtot \delta}{(4/3)\pi \rho_{\mathrm{d}} a^3}\cm3.
\end{equation}
The volumetric mass of solid is $1.386\ \mathrm{amu}\ \nHtot/\delta$, where $\delta$ is the dust to gas mass ratio assumed to be 0.01. The chemisorbed water over total solid mass ratio in percentage is
\begin{equation}
r_{\rm H_2O,\%}\equiv0.01\frac{3m_{\rm H_2O}(1+N_{\rm core,layer}) \Ns}{\rho_{\mathrm{d}} a}\simeq 0.045 \frac{1+N_{\rm core,layer}}{a^{\mu m}},
\label{eqn_core_layers}
\end{equation} 
which corresponds to $r_{\rm H_2O,\%} \sim$~2.7\% for 5 core layers (plus the top layer) and an average grain radius of 0.1 micron. It should be noticed that the number of core layers is not constrained in the model so that the actual parameter that governs the amount of chemisorbed water is $(1+N_{\rm core,layer})/a^{\mu m}$. In reality minerals have a maximum water storage capacity with the Ringwoodite mineral capable of storing up to $\sim$3\% of water in weight \citep{Kohlstedt1996}. This is consistent with the value of 2.7\% used in this study although storage capacities differ from a mineral to another one with capacities up to 6\% \citep{Hirschmann2005E&PSL.236..167H}. Alternatively, one can set the chemisorbed water over total solid mass ratio as the input parameter and derive the corresponding number of layers in the silicate core.
Therefore, we assume up to five layers in the silicate core can be hydrated (\Ncorelay~=~5).
The volume ratio of the layers of thickness $\Delta a$, with $a$ being the grain radius, to the total silicate core volume is $3 \Delta a/a$. Considering that a silicate layer thickness is 3 \AA, five layers plus the top layer correspond to 54 \AA, which gives $3 \Delta a/a = 54/333 \simeq 0.054$ or 5.4\%. We adopted a PAH abundance of 3 $\times$ 10$^{-7}$, i.e.\ $f_{\rm PAH}$~=~1 \citep{Tielens2005pcim.book.....T}.

\begin{table}
\begin{center}
\caption{Elemental abundances. \label{tab_elements}}
\vspace*{-0.5mm}          
\begin{tabular}{lc}     
\toprule
Elements & Abundance \\ 
\noalign{\smallskip}     
\hline
\noalign{\smallskip}  
H & 12.00\\
He & 10.984\\
C   & 8.14\\
N   & 7.90\\
O   & 8.48\\
Ne  & 7.95\\
Na  & 3.36\\
Mg  & 4.03\\
Ar    & 6.08\\
Fe   & 3.24\\
S    & 5.27\\
Si   & 4.24\\
PAH & see text\\
*       & pseudo-element, see text\\
!       & pseudo-element, see text\\
\noalign{\smallskip}
\bottomrule
\end{tabular}
\vspace*{-2mm} 
\tablefoot{The abundances are $10^{(x-12)}$ relative to H nuclei.}
\end{center}
\end{table}
\subsection{Simple zero-dimensional warm-up models}\label{mc_warmup_models}

The models are zero-dimensional pure chemical models where the physical conditions are set (density, gas and dust temperature, UV field, extinction, Cosmic-Ray flux, dust grain mean radius). We ran a typical cold molecular cloud model ($\nHtot$ =2$\times$10$^{4}$ $\cm3$, $\Td=\Tg=$~10~K) for 10 Myrs to set the initial abundances for the warm-up runs. The initial abundances for all species are in the single or doubly ionised form (H$^+$, He$^+$, C$^{++}$, O$^{++}$, N$^{++}$, S$^{++}$, Mg$^{++}$, Na$^{++}$, Fe$^{++}$, Ne$^{++}$, Ar$^{++}$, PAH$^{3+}$) for both the zero-dimensional and disk model. 
The instantaneous warm-up runs consist of models with increasing temperatures (still assuming $\Td=\Tg$) from 20 to 900~K. The silicate is assumed "dry" at the beginning of the chemical run.
We monitor in particular the abundance of water molecules in its different forms: gas-phase, \physi ice, chemisorbed water on the silicate surface, in the silicate core. The elemental abundances are listed in Table~\ref{tab_elements}. All the species modelled in this study are shown in Table~\ref{tab:standard-species}.
It should be stressed that the zero-dimensional model has been chosen to illustrate the effects of increasing gas and dust temperature on the formation of \phyllos\  and does not attempt to simulate an actual astrophysical environment.
\begin{table*}[!htbp]
\caption{Gas and solid species in the network.}
\label{tab:standard-species}
\vspace*{1mm} 
\begin{tabular}{cp{11cm}r}
\toprule
chemical elements & H, He, C, N, O, Ne, Na, Mg, Si, S, Ar, Fe & 12 \\
pseudo elements & *, ! & 2 \\
\noalign{\smallskip}
\hline
\noalign{\smallskip}
(H)         & H, H$^+$, H$^-$, H$_2$, H$_2^+$, H$_3^+$, H$_2^{\rm exc}$          &  7 \\
(He)        & He, He$^+$,                                                          &  2 \\
(C-H)       & C, C$^+$, C$^{++}$, CH, CH$^+$, CH$_2$, CH$_2^+$ 
              CH$_3$, CH$_3^+$, CH$_4$, CH$_4^+$, CH$_5^+$                     & 12 \\
(C-N)       & CN, CN$^+$, HCN, HCN$^+$, HCNH$^+$                               & 5 \\  
(C-O)       & CO, CO$^+$, HCO, HCO$^+$, CO$_2$, CO$_2^+$, HCO$_2^+$         & 7 \\
(N-H)       & N, N$^+$, N$^{++}$, NH, NH$^+$, NH$_2$, NH$_2^+$, NH$_3$, NH$_3^+$, NH$_4^+$   & 10 \\
(N-N)       & N$_2$, N$_2^+$, HN$_2^+$                                                   & 3 \\
(N-O)       & NO, NO$^+$                                                             & 2 \\  
(O-H)       & O, O$^+$, O$^{++}$, OH, OH$^+$, H$_2$O, H$_2$O$^+$, H$_3$O$^+$     & 8 \\
(O-O)       & O$_2$, O$_2^+$                                                  & 2 \\
(O-S)       & SO, SO$^+$, SO$_2$, SO$_2^+$, HSO$_2^+$                          & 5 \\
(S-H)       & S, S$^+$, S$^{++}$                                                  & 3 \\
(Si-H)      & Si, Si$^+$, Si$^{++}$, SiH, SiH$^+$, SiH$_2^+$                     & 6 \\    
(Si-O)      & SiO, SiO$^+$, SiOH$^+$                                    & 3 \\    
(Na)        & Na, Na$^+$, Na$^{++}$                                          &  3 \\
(Mg)        & Mg, Mg$^+$, Mg$^{++}$                                          &  3 \\
(Fe)        & Fe, Fe$^+$, Fe$^{++}$                                          &  3 \\
(Ne)        & Ne, Ne$^+$, Ne$^{++}$                                           &  3 \\
(Ar)        & Ar, Ar$^+$, Ar$^{++}$                                           &  3 \\
ice         & CO\#, H$_2$O\#, CO$_2$\#, CH$_4$\#, NH$_3$\#, SiO\#, SO$_2$\#, O$_2$\#, HCN\#, N$_2$\#  & 10 \\
\noalign{\smallskip}
\hline
\noalign{\smallskip}
pseudo species & *, ! & 2 \\
additional species & H\#, O\#, OH\#, *H\#, *\H2O\#, !\H2O\#, PAH, PAH$^-$, PAH$^+$, PAH$^{2+}$, PAH$^{3+}$, PAH\#, HPAH, HPAH$^+$, HPAH\#  & 15\\
\noalign{\smallskip}
\hline
\noalign{\smallskip}
 species    & total                                                           & 117 \\
\toprule
\end{tabular}
\end{table*}

\subsection{Hydration timescale models}\label{timescales}

We ran two series of models to determine the timescales as function of the temperature (assuming $T_{\mathrm{gas}}$=$T_{\mathrm{dust}}$) required to hydrate the silicate grains such that their mass percentages of water reach $\sim$ 2.7 \%. In the first series, the grain radius is fixed at 0.1 $\mu$m and the gas density is 2$\times$10$^4$ cm$^{-3}$, 2$\times$10$^8$ cm$^{-3}$, and 2$\times$10$^{12}$ cm$^{-3}$. Using formula~\ref{eqn_core_layers}, we derived the respective number of core layers to be 5, 58, and 60000 in addition to the top surface layer.  In the second series, we fixed the gas density at 10$^8$ cm$^{-3}$ and varied the grain radius (0.1 $\mu$m, 1 $\mu$m, 1 mm).

\subsection{{\sc DIANA-ProDiMo} \pd\ model}\label{standard model}
\begin{table}
\begin{center}
\caption{Cloud model parameters. \label{tab_cloud_models}}
\vspace*{-0.5mm}          
\resizebox{88mm}{!}{\begin{tabular}{lcc}     
\toprule
Parameter & symbol & values\\ 
\noalign{\smallskip}     
\hline
\noalign{\smallskip}  
Gas density & $\nHtot$        &   2$\times$10$^{4}$ $\cm3$\\
Temperature & $\Td=\Tg$     &  10, and 20 to 900~K\\
Extinction & A$_{\rm V}$             & 10\\ 
Strength of interstellar UV       & $\chi^{\rm ISM}$   & 1\\
Cosmic ray H$_2$ ionisation rate  & $\zeta_{\rm CR}$    & $\!\!1.7\times 10^{-17}$~s$^{-1}\!\!\!$\\
Grain radius & a   & 10$^{-5}$ cm\\
Dust-to-gas mass ratio & $\delta$ &  0.01\\
PAH abundance rel. to ISM         & $f_{\rm PAH}$         & 0.01, 1\\
Number of chemisorbed layers    & \Ncorelay & 5 \\
\noalign{\smallskip}
\bottomrule
\end{tabular}}
\vspace*{-2mm} 
\tablefoot{$\chi^{\rm ISM}$~=~1 is the ISM Draine UV field. $f_{\rm PAH}$~=~1 corresponds to a PAH abundance of 3 $\times$ 10$^{-7}$ \citep{Tielens2005pcim.book.....T}.}
\end{center}
\end{table}
The input for the {\sc DIANA} disk model are summarised in Table~\ref{tab:refmodel}. A detailed explanation of the parameters can be found in \cite{Woitke2016A&A...586A.103W}. The disk model uses a dust size distribution instead of a single grain radius. Dust grains can settle in the disk as parametrised by the turbulent mixing parameter $\alpha_{\rm settle}$. We ran the model in the fixed vertical structure mode where the both the radial and vertical density structures are parametrised. The chemical network is very limited with the addition of simple surface species and the surface chemisorbed species (for H and \H2O) and core chemisorbed species (\H2O only). The base network is described in \cite{Kamp2017A&A...607A..41K}. Most of the rates are taken from the UMIST2012 database \citep{McElroy2013A&A...550A..36M}. Contrary to the zero-dimensional model, photodesorption for the \physi species is accounted for.  Photodesorption does not occur for chemisorbed species. The UV field in the \pd\ is computed by solving the continuum radiative transfer \citep{Woitke2009A&A...501..383W}. We set for the \pd\ model a maximum of 500 layers (\Ncorelay~=~500) in the silicate core such that the average maximum uptake of water in silicate is 2 to 3\% of the total mass in solid form not taking water ice into account.
\begin{table}
\begin{center}
\caption{Model parameters, and values for the reference model.}
\vspace*{-0.5mm}
\label{tab:refmodel}
\resizebox{88mm}{!}{\begin{tabular}{lcc}
\\[-3.8ex]
\toprule
 quantity & symbol & value\\
 \noalign{\smallskip}
\hline 
\noalign{\smallskip}
stellar mass                      & $M_{\star}$      & $0.7\,M_\odot$\\
effective temperature             & $T_{\star}$      & $4000\,$K\\
stellar luminosity                & $L_{\star}$      & $1\,L_\odot$\\
UV excess                         & $f_{\rm UV}$     & $0.01$\\
UV powerlaw index                 & $p_{\rm UV}$     & $1.3$\\
X-ray luminosity                  & $L_X$           & $10^{30}\rm erg/s$\\
X-ray emission temperature        & $T_{X,\rm fit}$  & $2\times10^7$\,K\\
\noalign{\smallskip}
\hline
\noalign{\smallskip}
strength of interstellar UV       & $\chi^{\rm ISM}$ & 1\\
strength of interstellar IR       & $\chi^{\rm ISM}_{\rm IR}$ & 0\\
cosmic ray H$_2$ ionisation rate  & $\zeta_{\rm CR}$   
                                  & $\!\!1.7\times 10^{-17}$~s$^{-1}\!\!\!$\\
\noalign{\smallskip}
\hline
\noalign{\smallskip}
disk mass                & $M_{\rm disk}$   & $0.01\,M_\odot$\\
dust/gas mass ratio      & $\delta$        & 0.01\\
inner disk radius                 & $R_{\rm in}$     & 0.07\,AU\\
tapering-off radius               & $R_{\rm tap}$    & 100\,AU\\
column density power index        & $\epsilon$      & 1\\
reference scale height         & $H_{\rm g}(100\,{\rm AU})$ & 10\,AU\\
flaring power index               & $\beta$         & 1.15\\ 
\noalign{\smallskip}
\hline
\noalign{\smallskip}
minimum dust particle radius      & $a_{\rm min}$         & $0.05\,\mu$m\\
maximum dust particle radius      & $a_{\rm max}$         & $3\,$mm\\
dust size dist.\ power index      & $a_{\rm pow}$         & 3.5\\
turbulent mixing parameter        & $\alpha_{\rm settle}$ & 0.01\\
max.\ hollow volume ratio         & $V_{\rm hollow}^{\rm max}$   & 80\%\\
dust composition$^{(1)}$                    & $\rm Mg_{0.7}Fe_{0.3}SiO_3$ & 60\%\\
(volume fractions)                & amorph.\,carbon            & 15\%\\
                                  & porosity                   & 25\%\\
\noalign{\smallskip}
\hline
\noalign{\smallskip}
PAH abundance rel. to ISM         & $f_{\rm PAH}$        & 0.01\\
chemical heating efficiency$^{(2)}$        & $\gamma^{\rm chem}$  & 0.2\\
\noalign{\smallskip}
\hline   
\noalign{\smallskip}
distance                          & $d$ & 140\,pc\\
disk inclination                  & $i$ & 45\degr\\
\noalign{\smallskip}
\hline   
\noalign{\smallskip}
photodesorption   & Yield & 10$^{-2}$--10$^{-3}$\\
Number of chemisorbed layers    & \Ncorelay & 500 \\
\bottomrule
\end{tabular}}
\end{center}
\resizebox{90mm}{!}{
\begin{minipage}{100mm}{$^{(1)}$ The dust composition and porosity best match \pd\ observations of \pd s, although the porosity of the comet \object{67P/Churyumov-Gerasimenko} is 75-85 per cent in volume \citep{Herique2016}. $^{(2)}$  The chemical
      heating efficiency $\gamma^{\rm chem}$ is an efficiency by which
      exothermic chemical reactions are assumed to heat the gas. A detailed discussion on the disk parameters and their effects on the disk thermal and chemical structure can be found in \citet{Woitke2016A&A...586A.103W}. The photodesorption yield depends on the species. Special rates are used for water ice \citep{Westley1995Natur.373..405W,Oberg2009ApJ...693.1209O} and CO ice \citep{Munoz2010A&A...522A.108M}. For the other ice species, a standard yield of 10$^{-3}$ is assumed.}
\end{minipage}}
\end{table}  

\section{Results}\label{results}
\begin{figure*}[!ht]
  \centering  
   \includegraphics[angle=0,width=9.0cm,height=7.5cm,trim=25 70  70 300, clip]{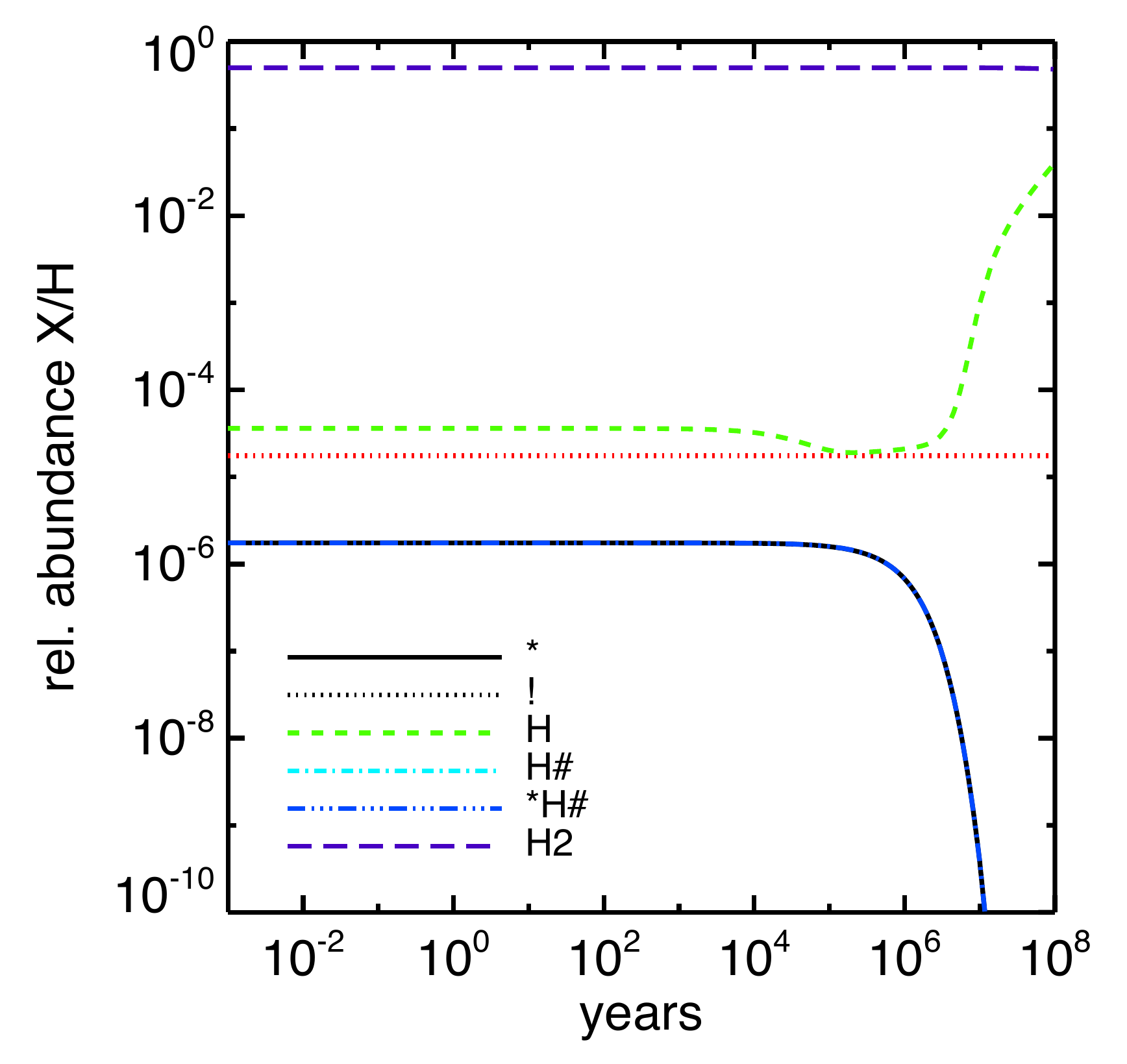}  
  \includegraphics[angle=0,width=9.0cm,height=7.5cm,trim=25 70  70 300, clip]{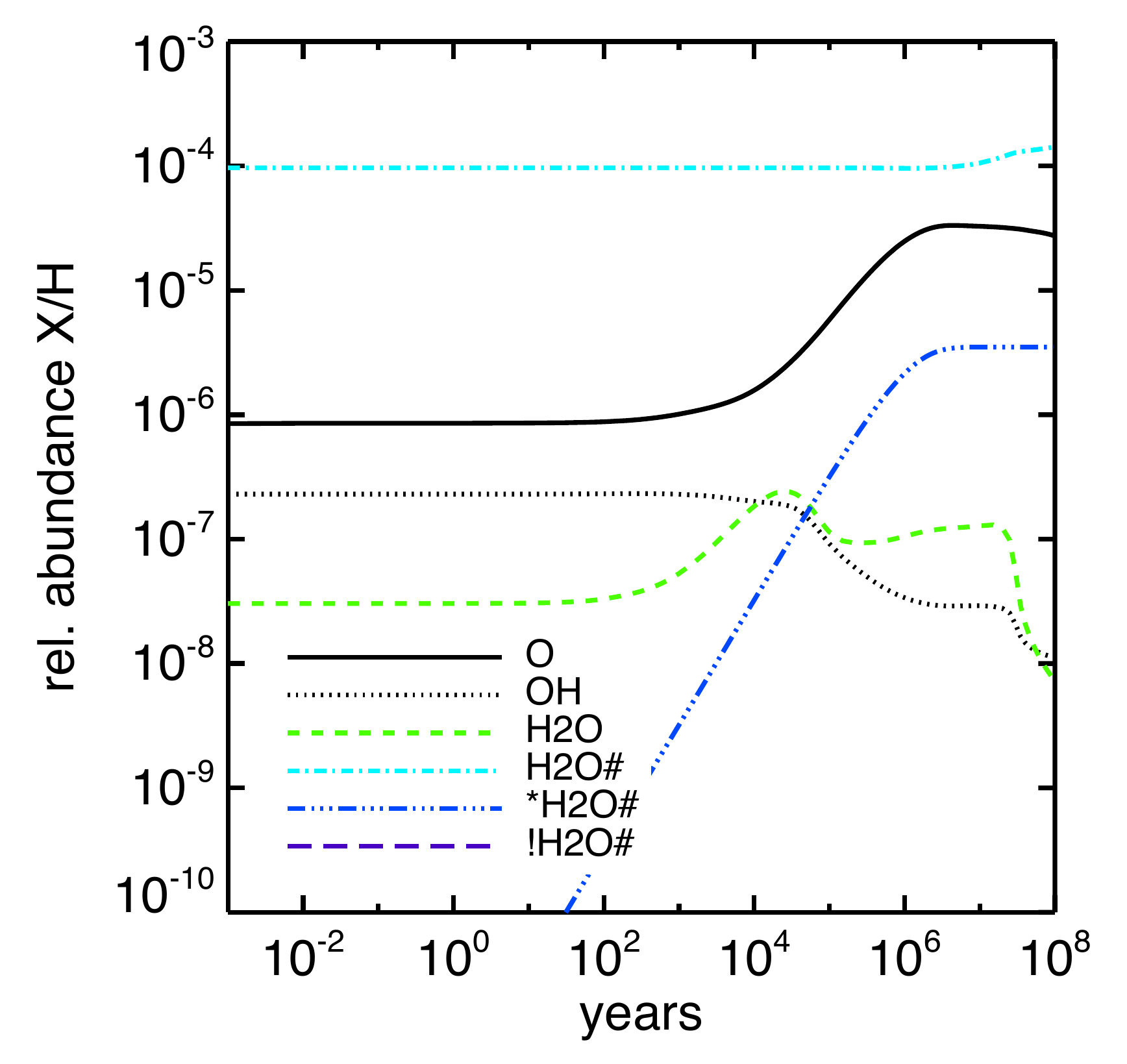}
   \includegraphics[angle=0,width=9.0cm,height=7.5cm,trim=25 70  70 300, clip]{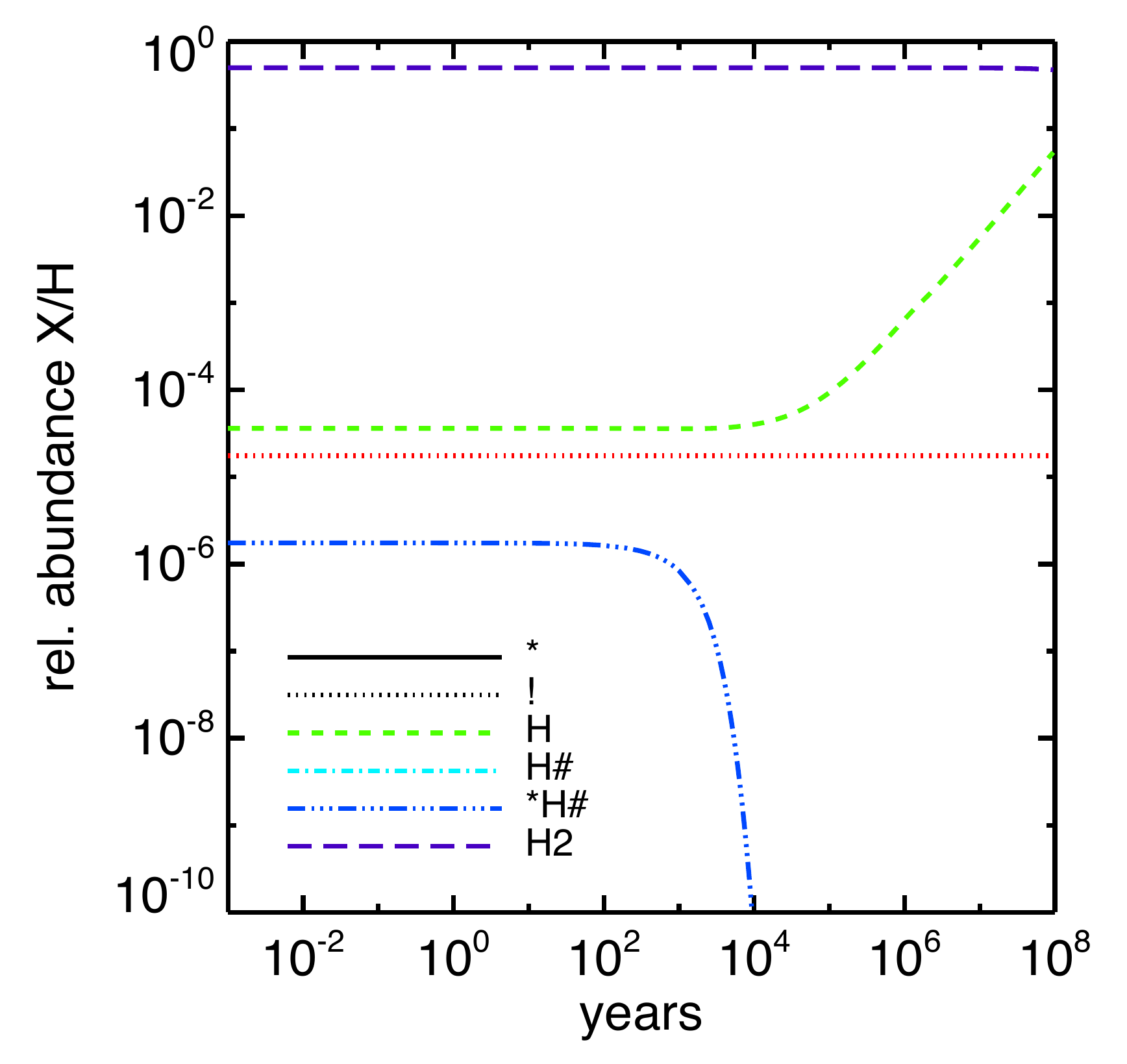}  
  \includegraphics[angle=0,width=9.0cm,height=7.5cm,trim=25 70  70 300, clip]{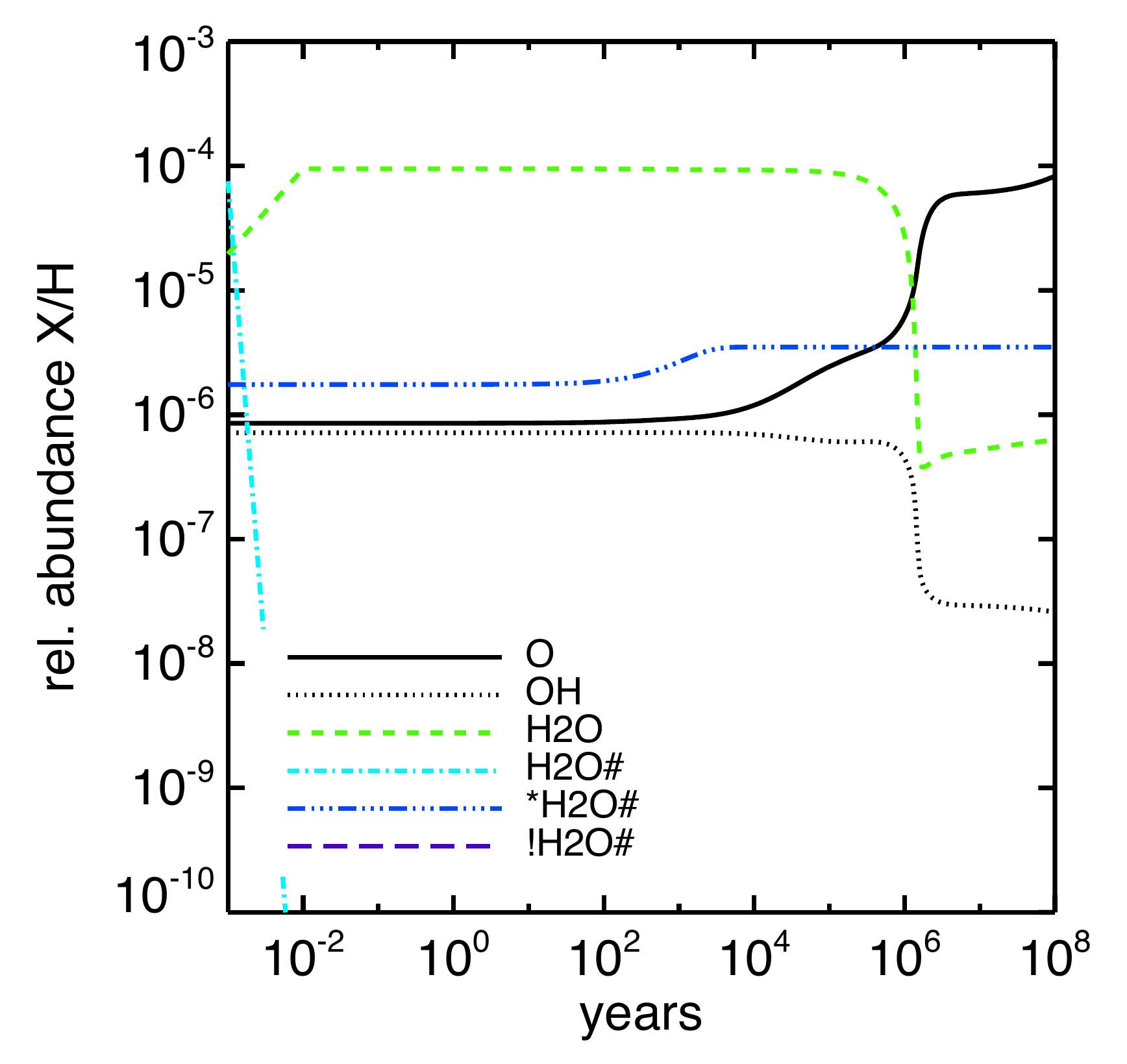}
   \includegraphics[angle=0,width=9.0cm,height=7.5cm,trim=25 70  70 300, clip]{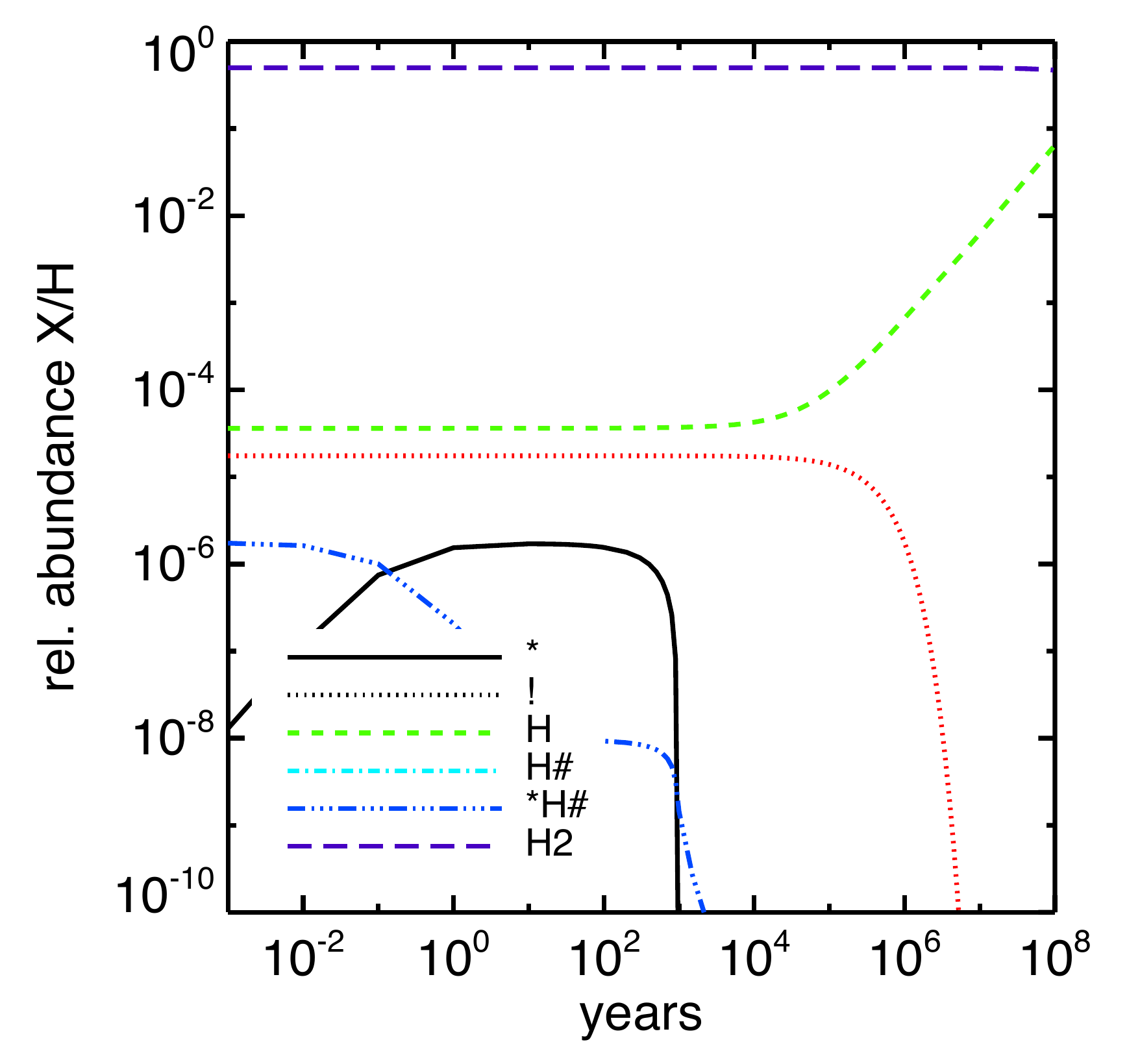}  
  \includegraphics[angle=0,width=9.0cm,height=7.5cm,trim=25 70  70 300, clip]{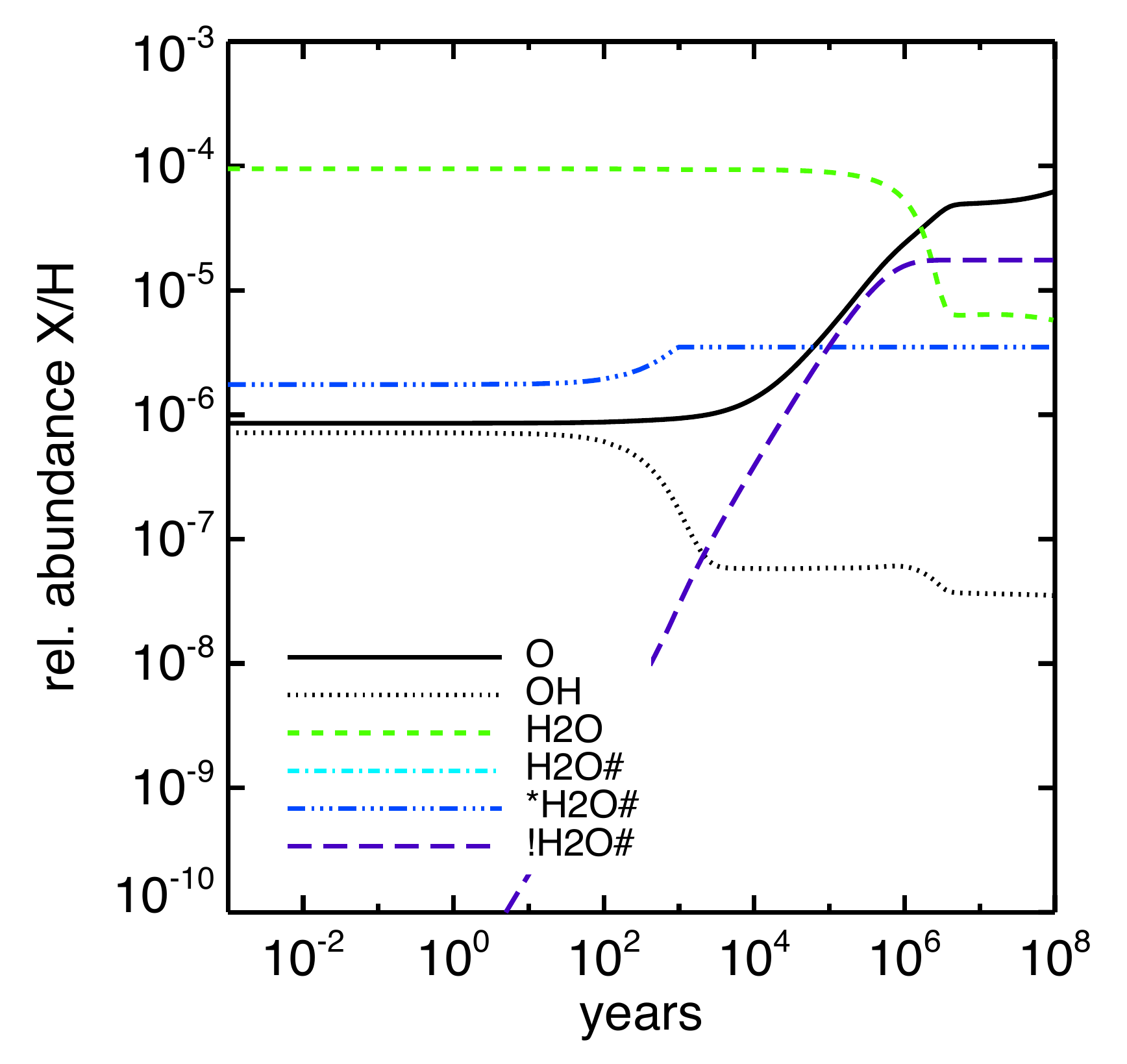}  
  \caption{Species abundances as function of time at 50~K (top panels), 150~K (middle panels), and 250~K (bottom panels). The abundance of unoccupied silicate surface (*) and core (!) sites, of atomic hydrogen in the gas and solid phases, and of \H2 are shown in the left panels. Gas-phase water and water ice abundances are shown on the right panels together with gas-phase abundance of atomic oxygen and OH.}
  \label{fig_water_abundances_time}          
\end{figure*}  

\subsection{zero-dimensional model results}\label{zeroD_models}

The results of models with warm-up temperature of 50, 150, and 250~K are shown in Fig.~\ref{fig_water_abundances_time}. The left panels contain the time variation of H, H chemisorbed on the grain silicate surface, \H2, water vapour, physisorbed \H2O\# and *\H2O\# chemisorbed on the grain silicate surface. The right panels are the abundance of water in the different reservoirs compared to the abundance of free surface  and core chemisorption sites.\\
\begin{figure*}[!ht]
  \centering
   \includegraphics[angle=0,width=9.0cm,height=7.5cm,trim=25 70  70 300, clip]{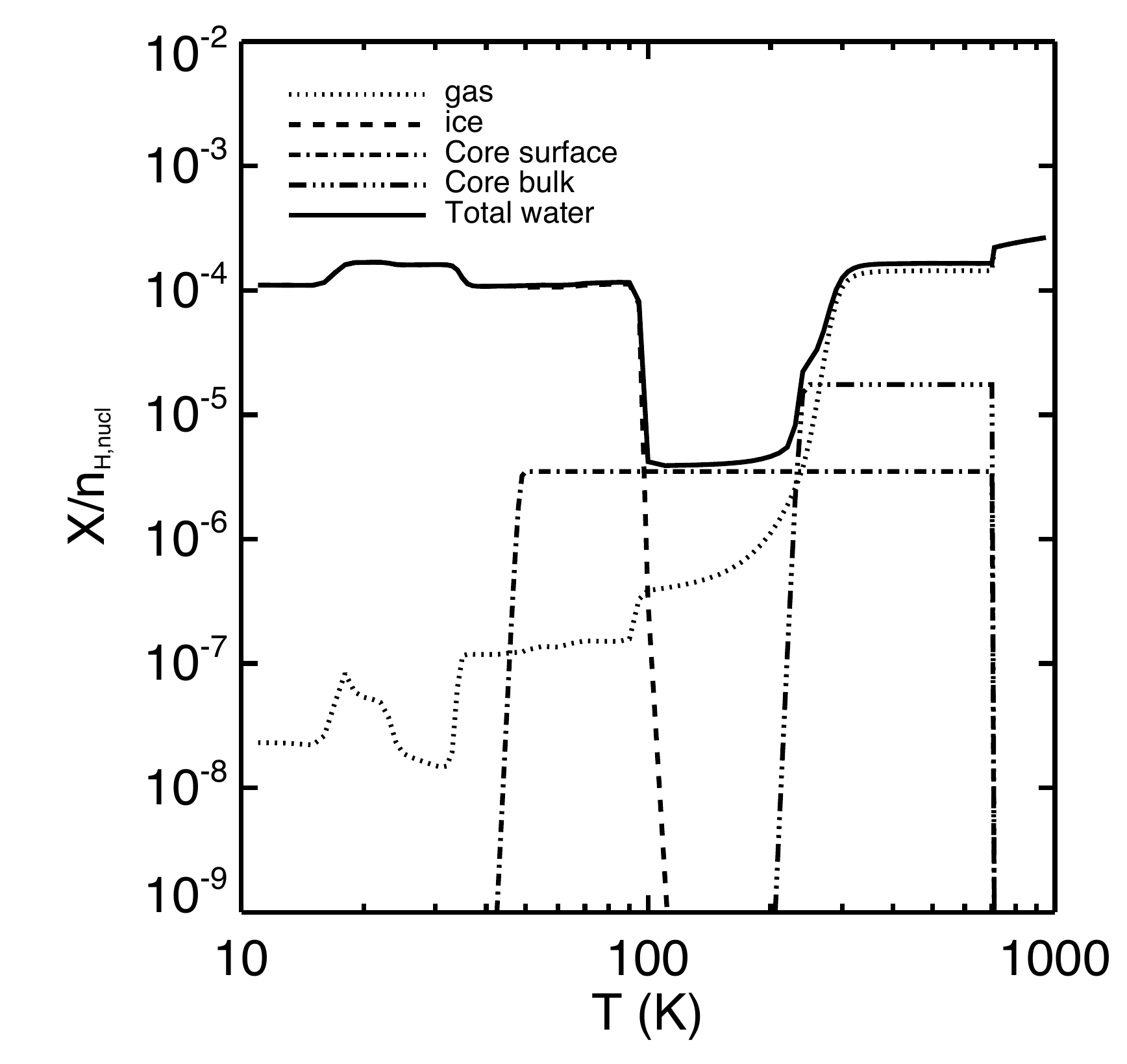}  
  \includegraphics[angle=0,width=9.0cm,height=7.5cm,trim=25 70  70 300, clip]{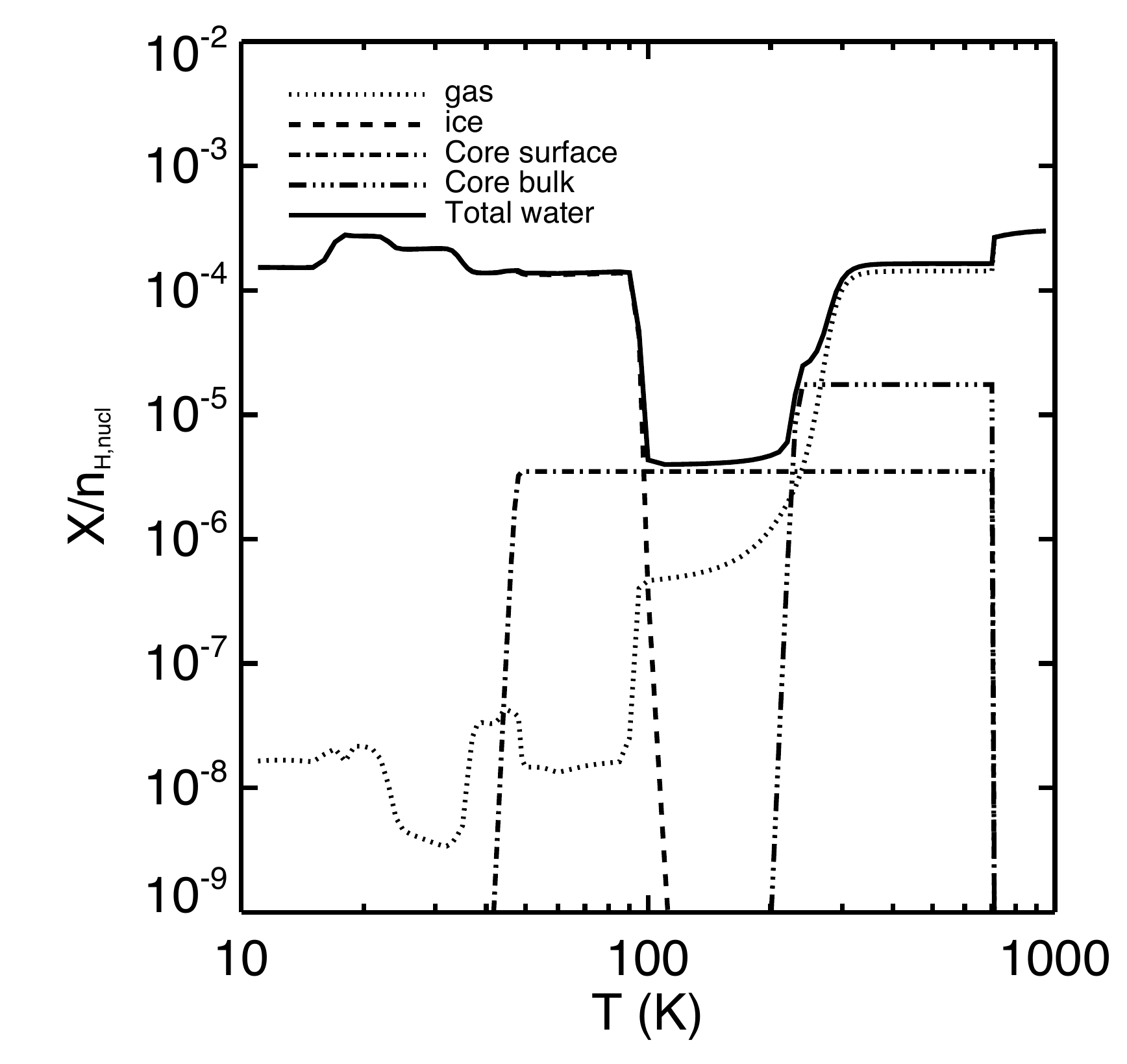}
  \caption{Water species abundances ($X/n_{\rm H,nucl}=X/\nHtot$) at 1 Myr (left panel) and 5 Myrs (right panel) as function of the gas and dust temperatures assuming $T_{\rm g}=T_{\rm d}=T$.}
  \label{fig_water_abundances_summary}          
\end{figure*}  
The abundance of \H2 stay relatively constant over time. On the contrary atomic hydrogen abundance increases as soon as all the chemisorption sites are occupied by water because \H2 is continuously destroyed and only reform at high dust temperature when H-atoms occupy the chemisorption sites. Once a water is bound to the silicate site, it is not reactive and can only diffuse to the surface if the site is in the silicate core and diffusion or desorb if the site is located at the silicate surface. The chemisorbed water prevent H-atoms to occupy the chemisorption sites.

Fig.~\ref{fig_water_abundances_summary} shows the different water reservoirs after 1 and 5 Myrs from the initial condition at 10~K. Below the desorption temperature of water ice at $\sim$~100~K for a gas at density 2$\times$10$^{4}$ cm$^{-3}$, most of the oxygen is in form of water ice (physisorbed water). 
The second reservoir up to $\sim$45~K with an relative abundance 10$^{-4}$--10$^{-3}$ times lower is water vapour. Gas-phase water abundance reaches a few 10$^{-8}$ consistent with detected abundance in molecular clouds \citep{Wirstrom2014ApJ...788L..32W}. The \phyllos\ formation activation energy is lower than the desorption energy of water at $\sim$5700~K.  Therefore, a surface water molecule can scan the entire surface and diffuse into the core before it desorbs as soon as the molecule has energy to overcome the barrier. In our density conditions, this happens from $\sim$45~K. The consequence is that the water molecules will occupy all the free chemisorption sites on the silicate surface. From 50 to $\sim$250~K, the molecules are not able to penetrate into the core because of the high activation barriers for diffusion but they occupied all the available surface chemisorption sites. The water abundance is the lowest between 100 and 200~K where only surface chemisorbed water is present. At temperatures between 100 and 250~K the gas-phase water formation  route is inhibited by the activation barriers \citep{Woitke2009A&A...501L...5W}. The slow gas-phase water formation cannot compensate the water destruction. Most of the oxygen is in form of atomic oxygen (see the middle panels in Fig.~\ref{fig_water_abundances_time}).
From 200~K, gas phase formation of water becomes again efficient to compensate for the decrease in \H2O abundance due to a lower \H2 abundance. From $\sim$100~K, all the surface chemisorption sites are occupied by water molecules. From $\sim$250~K, chemisorbed water molecules at the silicate surface start to penetrate in the core. From 250K till the desorption of water from chemisorption sites at $\sim$700~K all the chemisorption sites at the surface and in the silicate cores are occupied by water (-OH bonds) within 1 Myr. It should be noted that the desorption temperature depends on the species density $n_i$,  the thermal speed $v_i$, the surface density of site $\Ns$, and on the vibrational frequency of the species  in the surface potential well. Using the formula of \citep{Hollenbach2009ApJ...690.1497H} the desorption from chemisorption sites will occur at 847~K for a gas density of 10$^8$ cm$^{-3}$.
The abundances in the different water reservoirs change only slightly from 1 to 5 Myrs except for gas-phase water at low temperatures.
\begin{figure}[!ht]
  \centering  
   \includegraphics[angle=0,width=9.0cm,height=7.5cm]{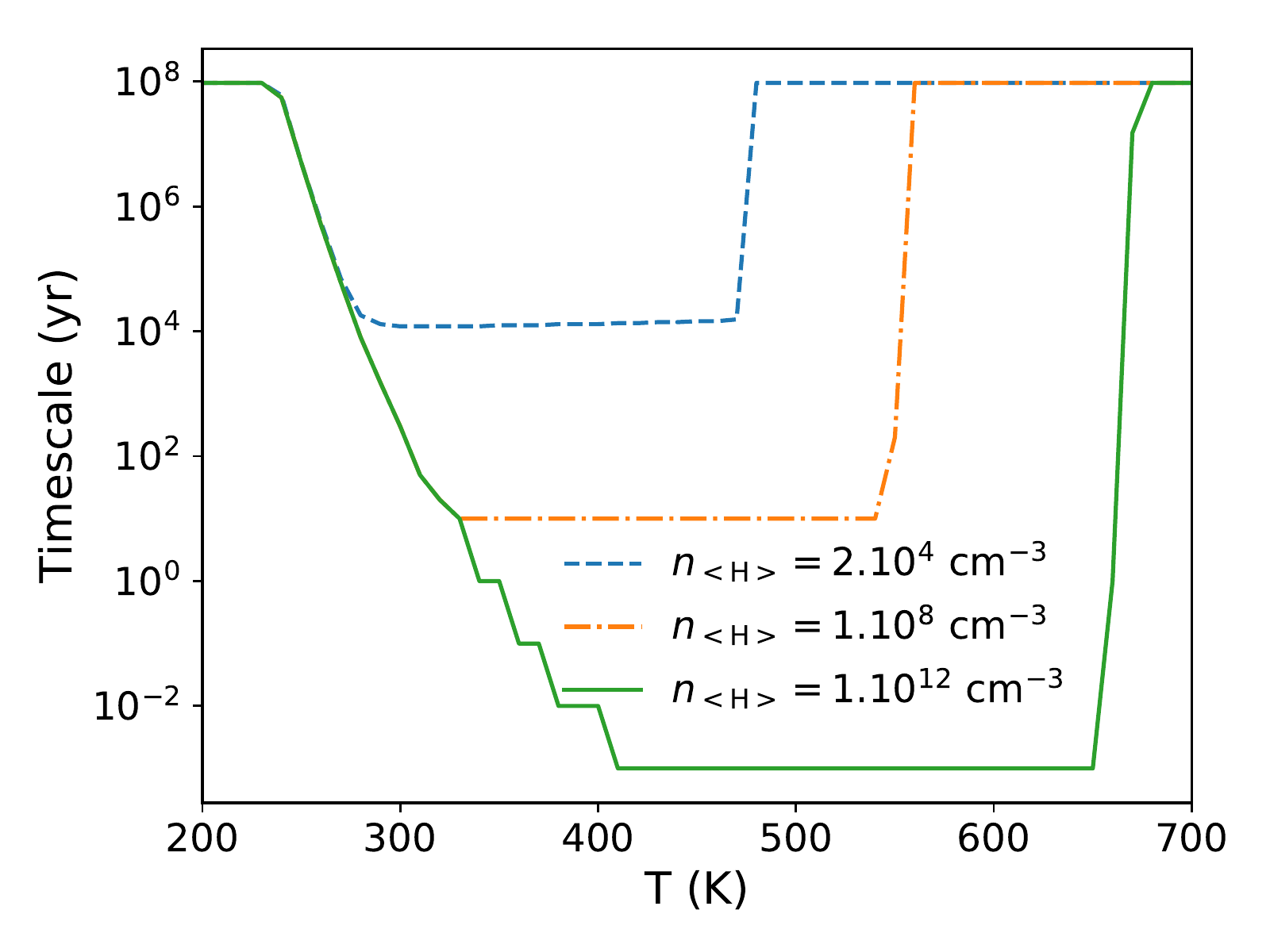}   
  \caption{Timescale for water vapor to adsorb to the 6 chemisorption layers of a  0.1 micron radius silicate grain. The timescales at 10$^8$ yrs mean that the water vapor  does not fill all the available silicate core layers within 10$^8$ years. }
\label{fig_water_core_timescale_density}          
\end{figure}  
\begin{figure}[!ht]
  \centering  
   \includegraphics[angle=0,width=9.0cm,height=7.5cm]{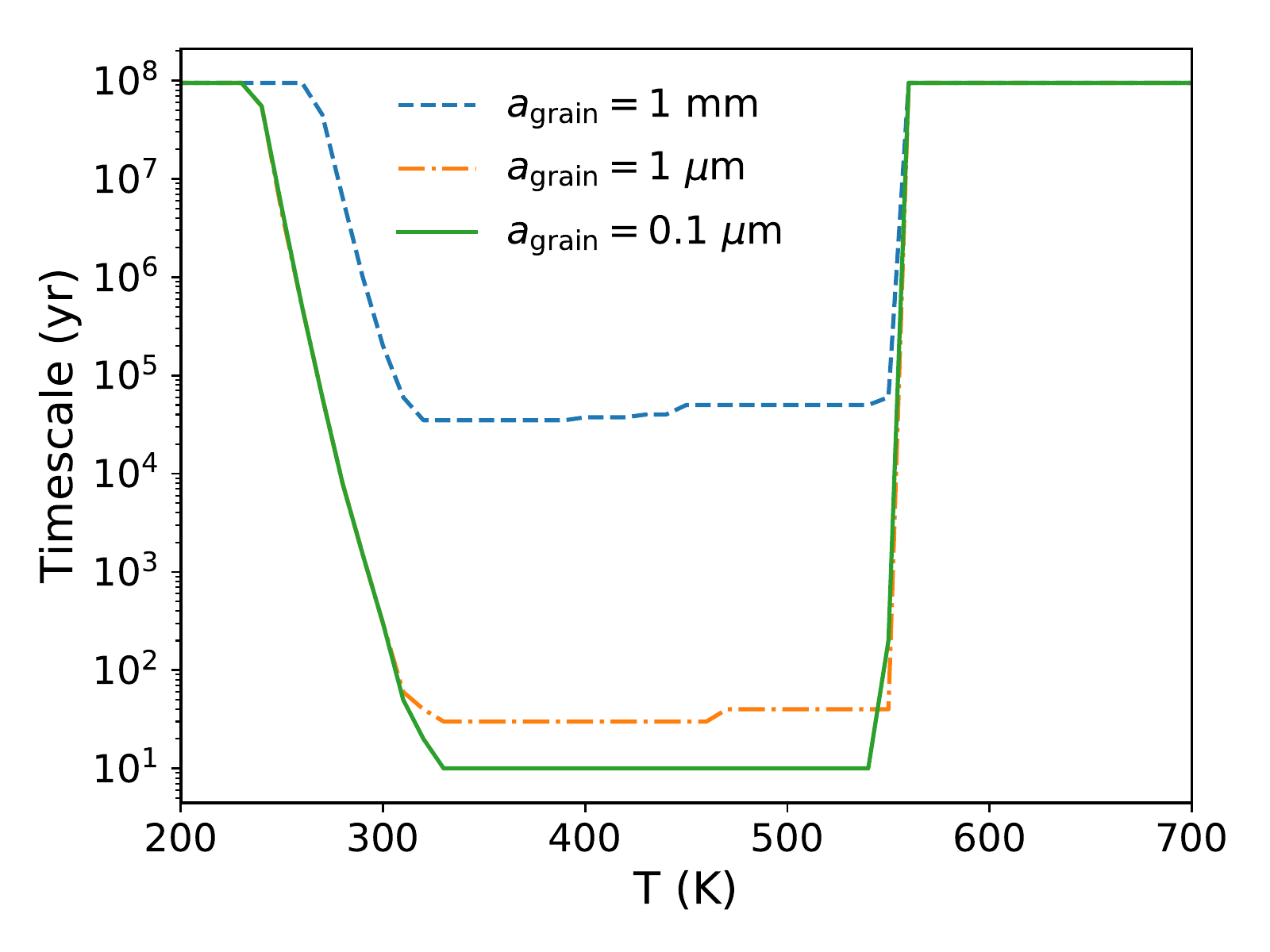}   
  \caption{Timescale for water vapor to make up $\sim$ 2.7 \% in mass of a silicate grain with radius 0.1 $\mu$m, 1 $\mu$m, and 1 mm with respectively 5, 58, and 60,000 silicate core layers. The gas density is $n_{\mathrm{<H>}}$= 10$^{8}$ cm$^{-3}$. The timescales are capped at 10$^8$ years.}
  \label{fig_water_core_timescale_large_grains}          
\end{figure}  

The reason of the ''blocking'' or ''poisoning'' of the chemisorption sites by water is that when two chemisorbed hydrogen atoms recombine, the unoccupied sites are quickly taken by a water molecule. Since in our model, chemisorbed water can only desorb and does not react, the chemisorbed sites are "poisoned" by water. The consequence is a significant decrease of \H2 formation by the recombination of two chemisorbed H atoms although most of the \H2 formation occurs via hydrogenated PAHs and PAH cations. This contrasts with \H2  formation in models without chemisorbed water where efficient \H2 formation takes place by encounters of chemisorbed H atoms (see Thi et al., submitted). All the chemisorption sites are rapidly occupied by water, the amount of chemisorbed water is only restricted by the maximum water uptake by silicates (see Sec.~\ref{discussion}).  When all the chemisorption sites are occupied by water molecules, the abundance of trapped water is still $\sim$~10 times less than the abundance of water vapour. Above 700~K, only water vapour remains with virtually all the available oxygen not in CO locked into water vapour. 

\subsection{Hydration timescales}\label{timescales_results}

The results of the models to determine the hydration timescales are shown in Fig.~\ref{fig_water_core_timescale_density} and Fig.~\ref{fig_water_core_timescale_large_grains}. The increase in gas density results in a higher desorption temperature. The hydration timescales $\delta t$ are below 10$^5$ years and vary with gas density and grain radius roughly as a/$n_{\mathrm{<H>}}$ at 450~K. The timescale dependence can be derived if one assumes that the hydration timescale is equal to the adsorption timescale : 
\begin{equation}
\left(4\pi a^2 \sqrt{\frac{2 k T}{m_{\mathrm{H_2O}}}}\epsilon(T) n_{\mathrm{H_2O}} n_{\mathrm{d}}\right)\delta t \simeq 4\pi a^2 N_{\mathrm{core,layer}}n_{\mathrm{d}},\label{eqn_timescale1}
\end{equation}
where $n_{\mathrm{H_2O}}$ is the number density of gas-phase water, $n_{\mathrm{d}}$ the number density of dust grains, $\sqrt{\frac{2 k T}{m_{\mathrm{H_2O}}}}$ is the gas thermal speed, and $\epsilon(T) \le 1$ is  the efficiency of the gas to overcome the adsorption and core diffusion activation barriers. Using formula \ref{eqn_core_layers}, we obtain for a fixed temperature
\begin{equation}
\delta t \ \propto\  r_{\rm H_2O} \left(\frac{a}{n_{\mathrm{H_2O}}}\right).
\end{equation}
At $T >$ 250K, the water gas-phase abundance reaches a maximum steady-state value of
a few 10$^{-4}$, such that
\begin{equation}
\delta t \ \propto\  r_{\rm H_2O} \left(\frac{a}{n_\mathrm{<H>}}\right).
\end{equation}
In the high density warm regions such as the inner disk midplane of protoplanetary disks, most of the silicate grains are hydrated up to their maximum possible water intake. The timescale increases with grain size so that the hydration timescale should be compared to that of  grain growth.

\subsection{Disk model results}\label{disks_models}
\begin{figure*}[!htbp]  
  \centering
  \includegraphics[angle=0,width=8.0cm,height=8cm,trim=50 80  80 300, clip]{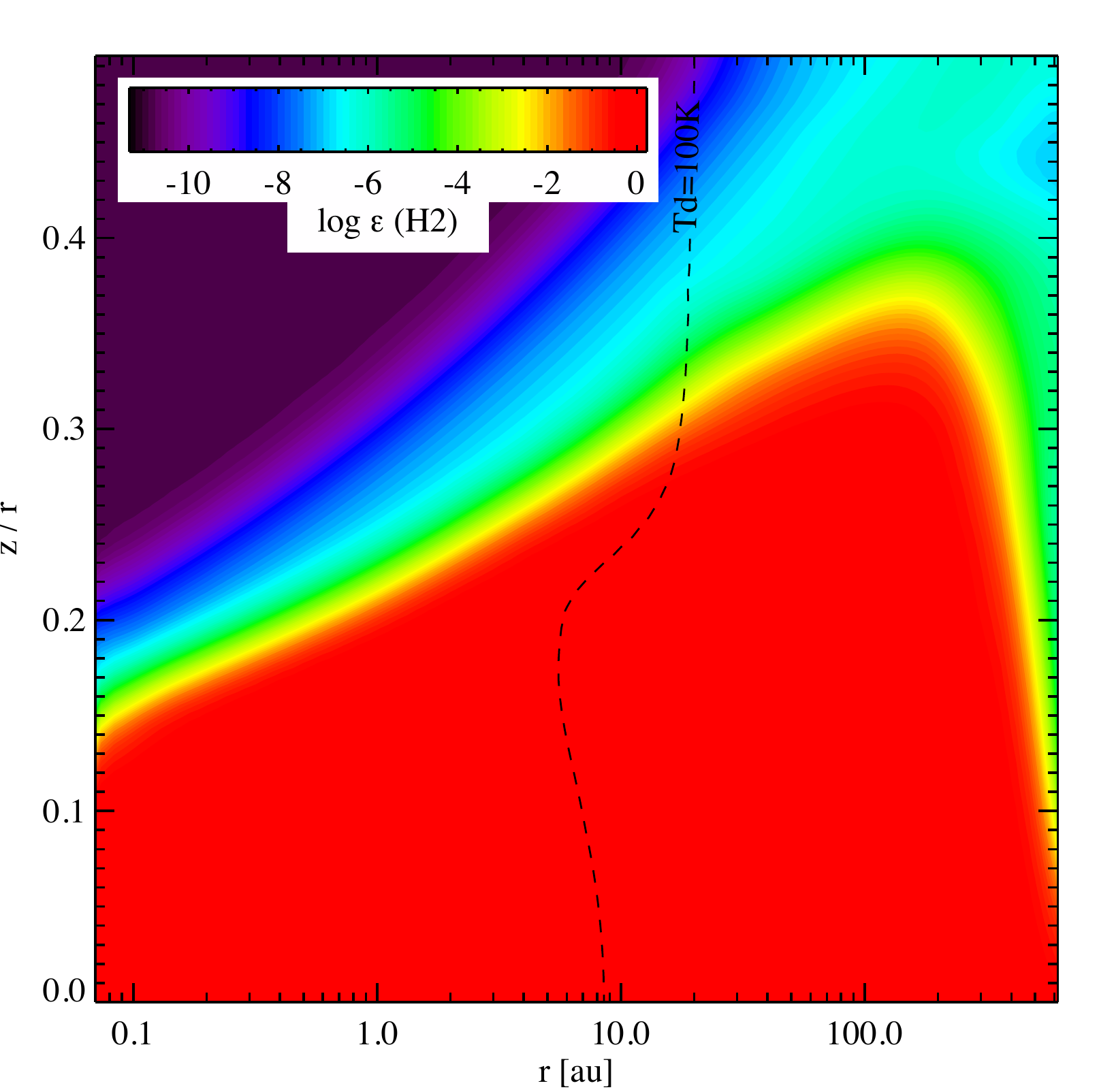}  
  \includegraphics[angle=0,width=8.0cm,height=8cm,trim=50 80  80 300, clip]{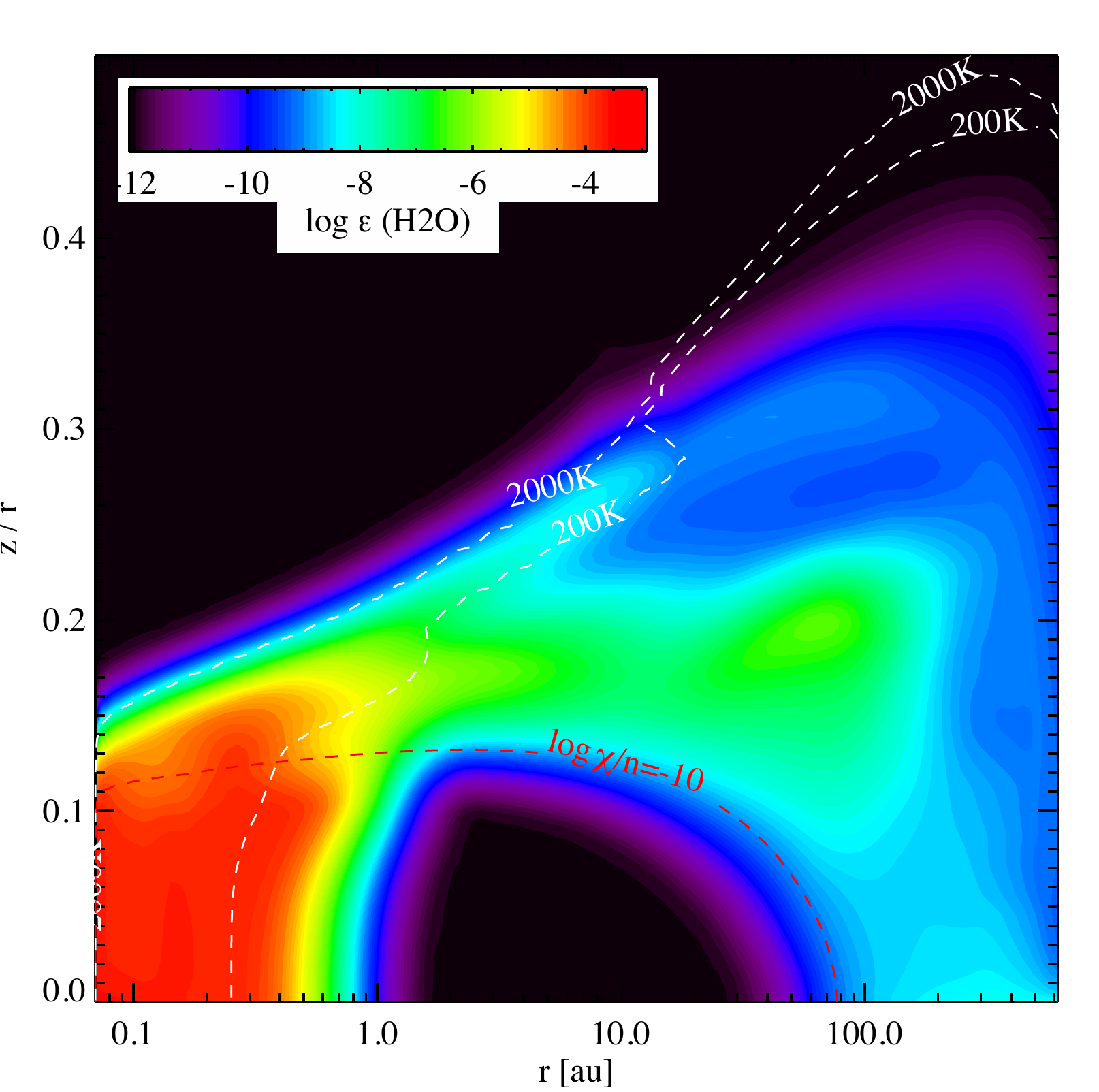}
  \includegraphics[angle=0,width=8.0cm,height=8cm,trim=50 80  80 300, clip]{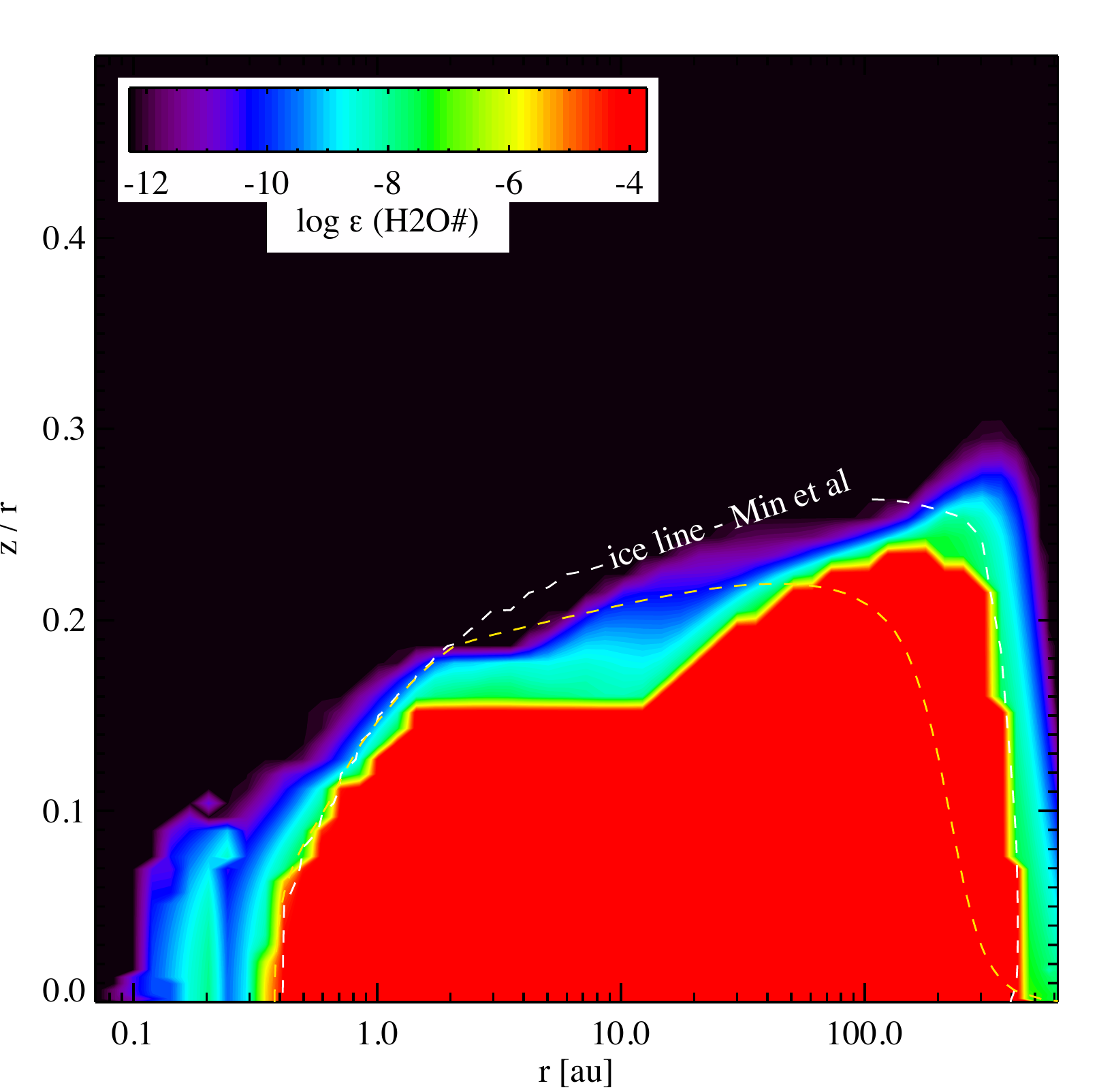}
  \includegraphics[angle=0,width=8.0cm,height=8cm,trim=50 80  80 300, clip]{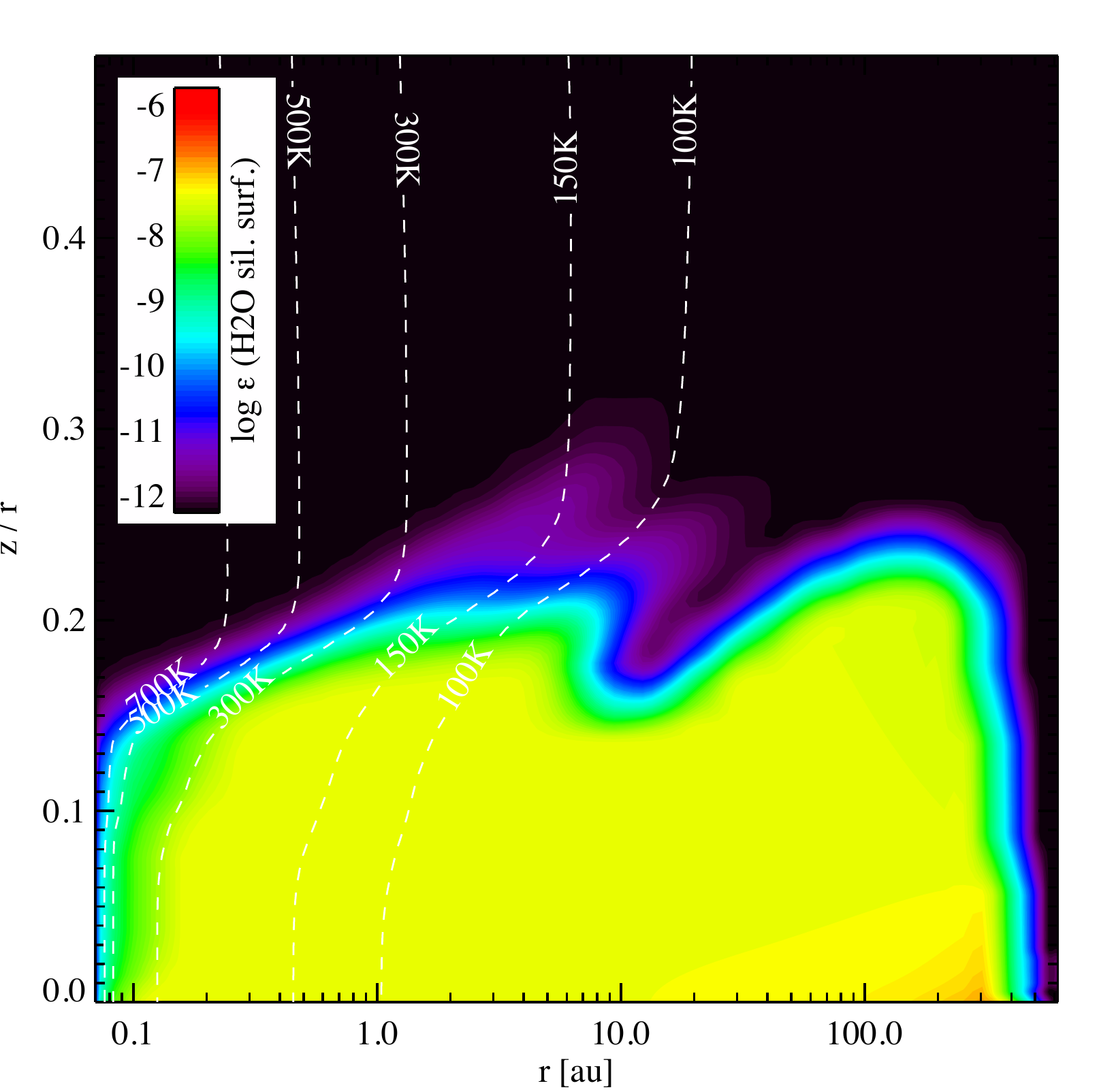}  
  \includegraphics[angle=0,width=8.0cm,height=8cm,trim=50 80  80 300, clip]{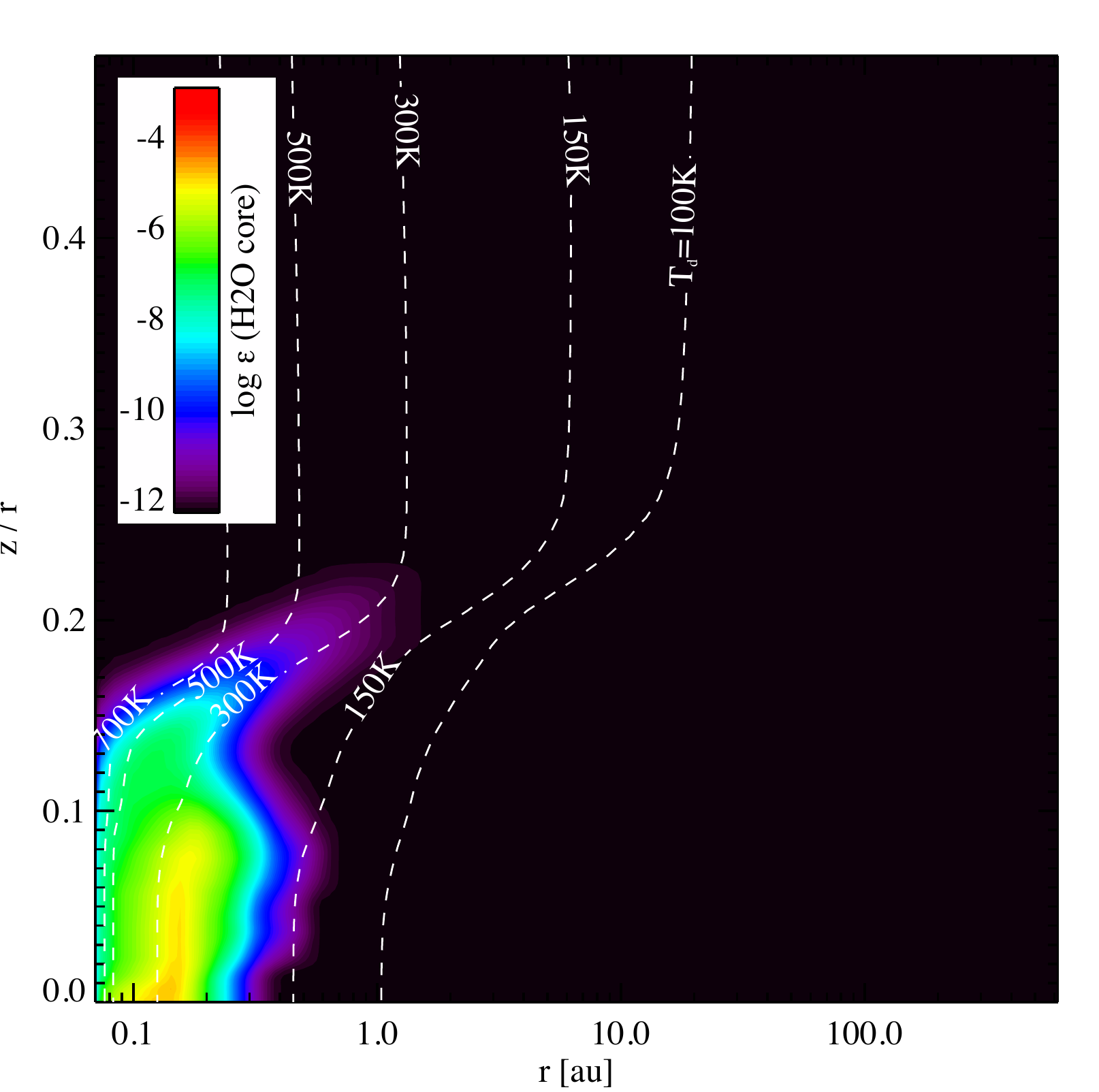}
  \includegraphics[angle=0,width=8.0cm,height=8cm,trim=50 80  80 300, clip]{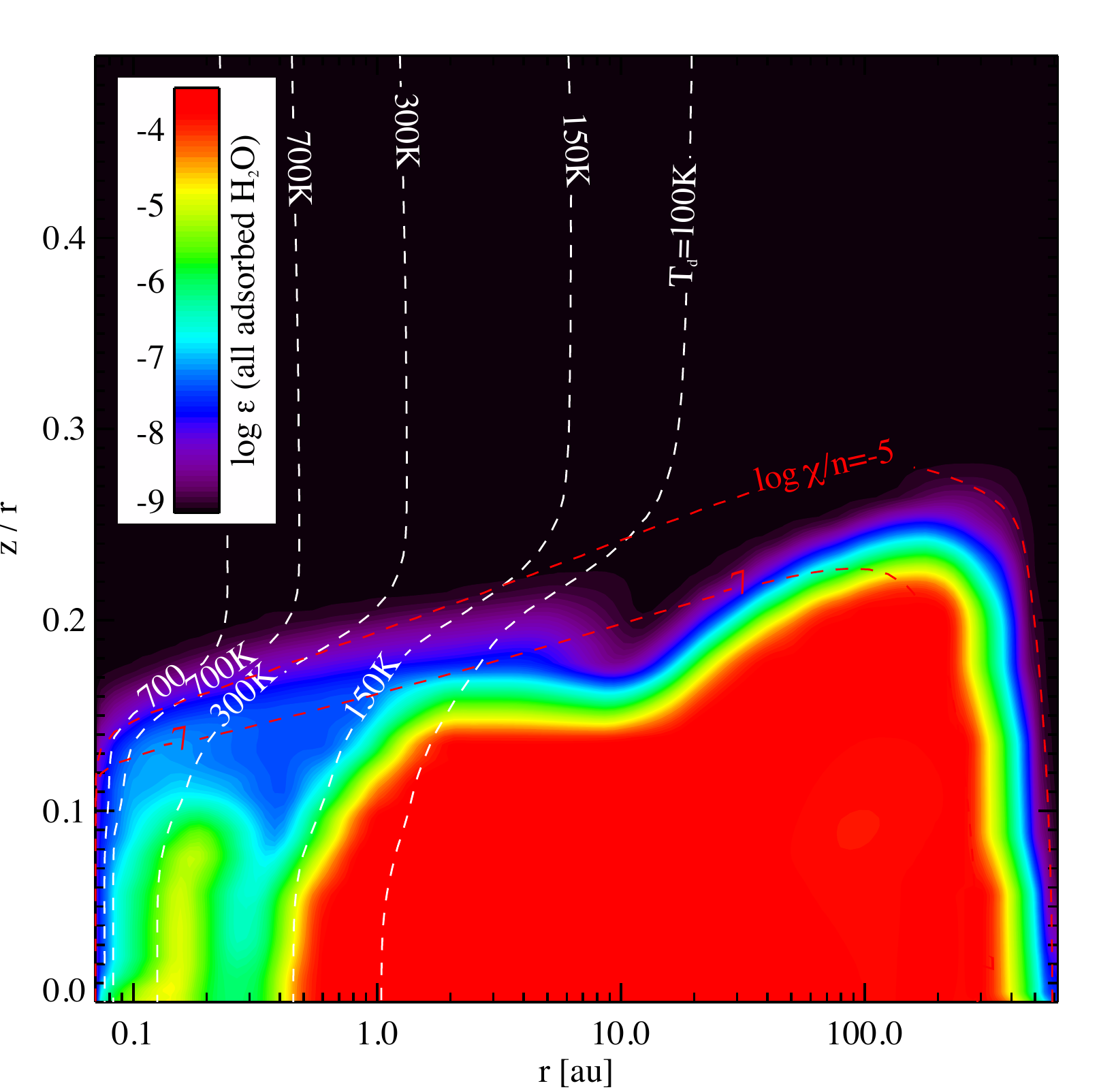}  
  \caption{Steady-state abundance for \H2 and the different \H2O carriers for the standard disk model (\H2O\# means \physi water). The white contours correspond to the gas temperature in the upper right panel. $\chi$ is the strength of the UV field ($\chi$~=~1 in average in the local interstellar medium. The water ice location analytical model of \citet{Min2016A&A...593A..11M} is shown. In the same panel the alternative formula from \cite{Antonellini2016A&A...585A..61A} is also displayed.}
  \label{fig_disk_results1}          
\end{figure*}  
\H2 and water abundances of the disk model with 500 core layers are shown in Fig.~\ref{fig_disk_results1}. The figure shows the location of the different water reservoirs in a typical (0.01 M$_\odot$)  \pd\ (gas-phase, water ice, chemisorbed water at the silicate core surface, chemisorbed water in the core, aka \phyllos). The abundance of water in all its possible adsorbed forms is shown in the lower-right panel of Fig.~\ref{fig_disk_results1}.\\
Hydrogen is almost entirely in molecular form except in the upper disk atmosphere layers in the outer disk (R$>$400 au). \H2 is formed from \physi H-atoms at gas and dust temperatures below 10~K and from chemisorbed H-atoms at high temperatures. A detailed discussion on the \H2 formation processes is given in Thi et al., in submitted.

At low temperatures as found in the outer disk ($\Td<$~100~K), the main water reservoir is physisorbed water (water ice). This is the so-called water ice zone. Photodesorption of physisorbed species is effective in the outer disk when $\chi /n >$ 10$^{-7}$ ($\chi$ being the enhancement compared to the standard interstellar UV field) since we see that the water ice zone does not extend further than $\sim$~400 au while the disk outer radius is $\sim$ 600 au. 

Within the inner 10 au where the dust is warmer than $\sim$~150~K, water vapour is the main oxygen reservoir. Water vapour is also abundant in the "atmosphere finger" \citep{Woitke2009A&A...501L...5W,Du2014ApJ...792....2D}. The gas-phase water chemistry can be complex and has been discussed in previous studies \citep{Kamp2013A&A...559A..24K,Thi2005A&A...438..557T,vanDishoeck2014prpl.conf..835V}. In the inner disk, both water on the silicate surface and in the core are present with the water being trapped in the silicate core only when the dust temperature is high enough for the water molecules to overcome the activation energy for diffusion into the silicate core.  Photodesorption is efficient in the inner disk upper atmospheres (see low-right panel of Fig.~\ref{fig_disk_results1} when $\chi /n >$ 10$^{-7}$) by restricting the lifetime of \physi water on the bare grain surface. Therefore, it also limits the efficiency of the transfer of \physi water to chemisorption water and subsequently influences the abundance of trapped water in the silicate core. Since the dust temperature even at 0.1 au is lower than 700~K, chemisorbed water can remain on the grains. In the \pd\ mid-plane regions where $\Td$ is between the sublimation of water ice and the efficient formation of \phyllos\ (9 to 20 au), the amount of water both in the gas and in the solid phases is low, consistent with the zero-dimensional model results (see Fig.~\ref{fig_water_abundances_summary}).
However, the abundance of chemisorbed water can reach its maximum possible value for dust grains hotter than $\sim$250~K (Fig.~\ref{fig_water_abundances_summary}).

Similar to the zero-dimensional models chemisorbed water acts as a "poison" limiting the amount of chemisorbed H atoms. The consequence is a lowering of the \H2 formation efficiency. The lower \H2 formation on silicate grains is however compensated in part by an increase in \H2 formation via hydrogenated PAHs and PAH cations \citep{Andrews2016AA...595A..23A,Boschman2015AA...579A..72B,Mennella2012ApJ...745L...2M}. The efficiency of \H2 formation through hydrogenated PAHs depends on the actual abundance of PAHs, which is so far not well determined in \pd s. In our disk model, the PAH abundance is set to be 100 times smaller than in the general interstellar medium ($f_{\rm PAH}$=0.01). The occupancy of chemisorption sites by water and the drop in \H2 formation efficiency is a side effect of the water chemisorption model. 

\section{Discussion}\label{discussion}

As seen in both the zero-dimensional and the \pd\ models, hydration of the anhydrous silicate into \phyllos\ is relatively efficient at temperatures above 250--300~K. Phyllosilicates are also the thermodynamically most stable forms of silicates in an oxygen-rich environment at 200-400~K \citep{Woitke2018A&A...614A...1W} and appendix \ref{lizardite}.

Grain growth by coagulation occurs rapidly and millimetre-sized grains are detected in \pd s \citep{Tazzari2016A&A...588A..53T}. The units making up the large grains can be the 0.1 micron grains considered in the zero-dimensional models. However, compact millimeter-size grains can be hydrated within 10$^5$ years if the gas density is above 10$^8$ cm$^{-3}$. Our results suggest that the maximum water storage capacity of silicates is easily reached within 1 Myr for gas hotter than $\sim$~250~K for gas density down to 2$\times$10$^{4}$ cm$^{-3}$. 

The speed of the \phyllos\ formation depends strongly on the activation energy. Previous works have used energy of $\sim$8420~K \citep{Fegley2000}, which is much higher than the current accepted value around 3000~K.
Interestingly such value for the activation energy is lower than the desorption energy for \physi water. This results in a transfer of water ice to the silicate core surface via tunnelling already at $\Td \sim$50~K. With an activation energy higher than the desorption energy, \phyllos\ would only be formed through the \ER\ process. Another difference with previous works is that at temperatures $\Td>$100~K water can still hop a few times from a physisorption site to another physisorption site and find an unoccupied chemisorption site before it desorbs back to the gas phase (\LH\ process). When the water molecule arrives in a chemisorbed site, a dust temperature of more than 700~K is required for it to evaporate. It should be noted that in the detailed Monte-Carlo simulation of \cite{DAngelo2018arXiv180806183D}, the silicate surface coverage by chemisorbed water is still 30\%.
\cite{Fegley2000} only considered hydration of silicates at the surface by the direct formation of a chemisorption bond, i.e.\ they only accounted for the thermal \ER\ process and not the precursor-mediated process. In addition to the thermal \ER\ process the gas-phase \H2O can also tunnel through the barrier although the effectiveness is small because of the high weight of the water molecule.

Many factors may limit the efficiency of embedding water in silicate grain cores. In the disk model, photodesorption competes with diffusion of the water molecule from a physisorption site to a chemisorption site even in the dust extinct inner disk region. In our models, we assume that the silicate surface is in contact with the icy mantle or faces directly the vacuum. If carbonaceous material covers the silicate core, then the water molecules have to diffuse through the carbonaceous layer before reaching the silicate core \citep[e.g.,][]{jones2017A&A...602A..46J}. 

As an alternative, direct formation of \phyllos\  can also occur by implementation of energetic H$^+$ \citep{Djouadi2011A&A...531A..96D}.

\begin{figure}[!ht]
  \centering  
   \includegraphics[angle=0,width=9.0cm,height=7.5cm,trim=25 70  70 300, clip]{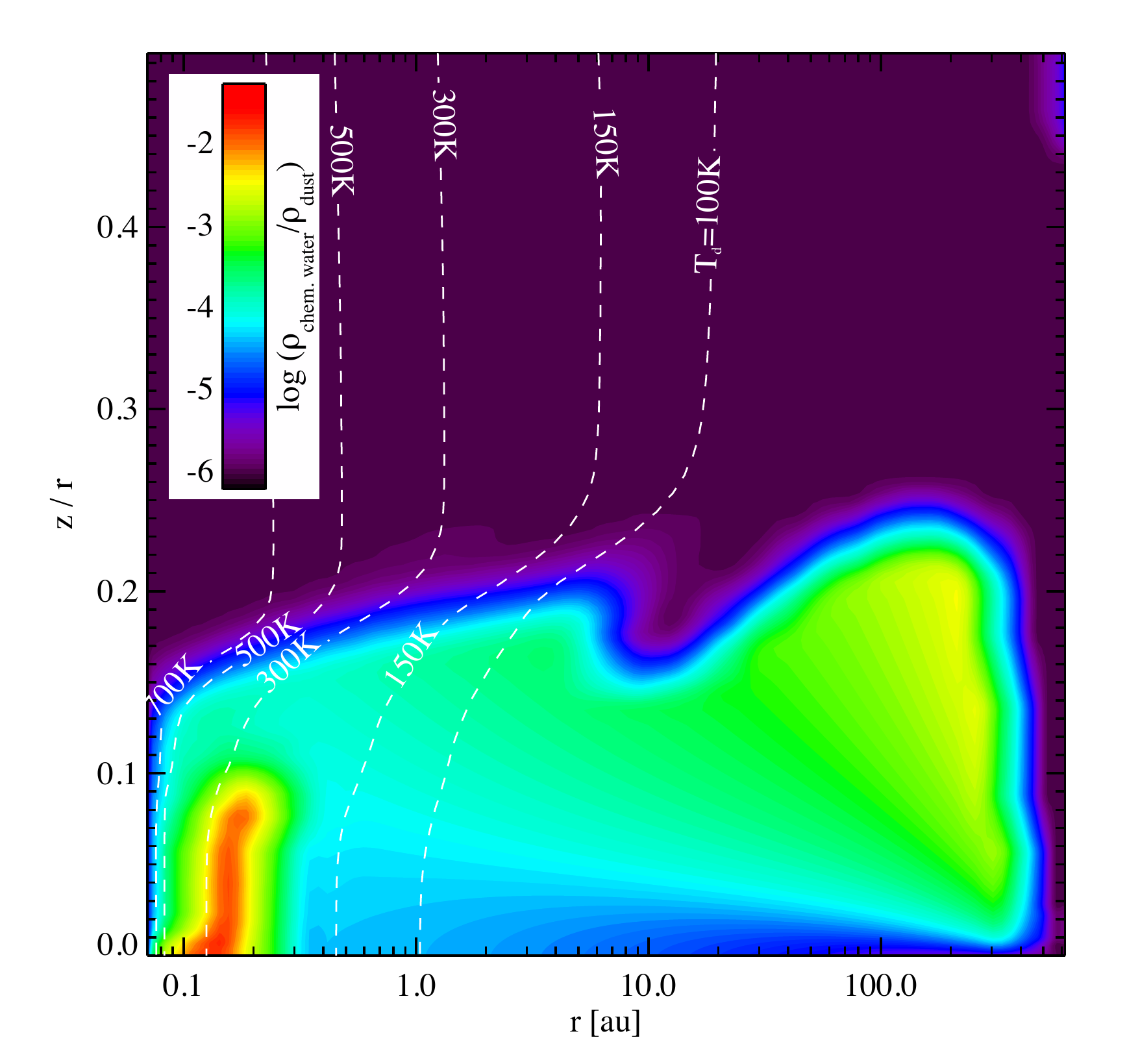}   
  \caption{Chemisorbed water mass over refractory dust mass ratio. The model assumes a maximum of 500 layers in the silicate core. The white contours correspond to the dust temperature computed from a 2D continuum radiative transfer.}
  \label{fig_water_dust_mass_ratio}          
\end{figure}

As \phyllos\ does not dehydrate before reaching 600-700~K (and at higher temperatures for dense gas in the inner disk regions), most of it will remain at the planet formation stage of \pd s. Fig.~\ref{fig_water_dust_mass_ratio} shows the ratio between water trapped into \phyllos\ form and refractory dust mass. One consequence of the fast \phyllos\ formation in contrast to formation by aqueous alteration is that \phyllos\ should be relatively uniformly abundant in partially differentiated planetesimals located in warm \pd\ regions (where $\Tg$ \& $\Td>$300~K), if the grain coagulation and growth of the planetesimals are slower than the uptake of water in the grains. 

Phyllosilicates can be detected through their OH-stretch fundamental absorption at about 2.7 $\mu$m. The infrared signature of \phyllos\ is ubiquitous at the surface of the dwarf planet Ceres \citep{Ammannitoaaf2016} and in a majority of other asteroids \citep{Takir2015}. Phyllosilicates also show distinctive features in the mid-infrared \citep{Glotch2007Icar..192..605G, Pitman2010_doi:10.1111}. Olivine crystal infrared features peak at 11.2 and 19.5 $\mu$m, while \phyllos\  have their maximum for the SiO$_4$ stretching at a lower wavelength (around 10 $\mu$m) in addition to  bands at 15.8 and 22 $\mu$m due to bending vibration of the hydroxyl group as found in meteorites \citep{Beck2014,GARENNE201493}. Amorphous olivine peaks at $\sim$~10 $\mu$m. The phyllosilicate features in meteorites differ in shape compared to terrestrial \phyllos s. 
The zodiacal dust mid-infrared spectrum suggests the presence of \phyllos s \citep{Reach2003Icar..164..384R}. \citet{Morris2009AsBio...9..965M} modelled the mid-IR emission of dust grains composed of anhydrous silicates and \phyllos s in a \pd, while \citet{Morlok2014Icar..231..338M} compared directly the IR spectra of \pd s with meteoritic samples. \cite{Min2016A&A...593A..11M} modelled in detailed the Spectral Energy Distribution of the \pd\ around \object{HD 142527} between 35 and 100 microns and could constrain the amount of \phyllos s to be lower than $\sim$47\% because the \phyllos\ are almost featureless in this wavelength range. Wavelength and shape variations with the level of crystallinity as well as effects of grain size render the identification of \phyllos s in space IR spectra difficult unless high signal-to-noise ratio spectra from the James Webb Space Telescope are available.

Phyllosilicates have been tentatively detected in a debris disk \citep{Currie2011}. Since the dust grains in a debris disk result from collisions between planetesimals, the origin of the \phyllos s cannot be easily established. Another potential test of the proposed \phyllos\ formation model is to observe that the average abundance of \phyllos\ is anti-correlated with that of water ice (see Fig.~\ref{fig_water_abundances_summary} and the lower right panel of Fig.~\ref{fig_disk_results1}). The water is either in the water ice or in the silicate core. Objects with a high water ice abundance should not show concomitantly a high \phyllos\ abundance unless there is extensive radial mixing at the early stage of planetesimal formation, which will bring dry silicates and \phyllos s in the outer disk, where they can be water ice coated \citep[e.g.,][]{Raymond2017}.

Chondritic fine-grained \phyllos s formation can be explained by our process. However, whether all observed \phyllos\ in meteorites and Solar System objects can be explained by our formation method is not clear. For instance the widespread presence of \phyllos\ at the surface of Ceres may be due to actual aqueous alteration at some stage in the object's geological history. Aqueous alteration should have occurred between 300 and 600~K \citep{Brearley2006}. This temperature is reached by the heat generated by the radioactive decay of the short-lived radionuclides Aluminium-26 \citep{GRIMM1989}. Formation of \phyllos\ can occur both in the pre-dust coagulation phase and in the planetesimals formation and evolution phase.  

Given the different possible sources of water on Earth, \cite{Izidoro2013ApJ...767...54I} proposed a compound model for the origin of Earth's water where they included all the possible sources.

The HDO/H$_2$O ratio in Earth's ocean water has been used as a constrained on the origin of the water on Earth \citep{Caselli2012A&ARv..20...56C,Drake2005M&PS...40..519D} . At face value the D/H ratio of the water formed in our model would be below Vienna Standard Mean Ocean Water (VSMOW) and even the values measured for the deep mantle.  Thus supply of high D/H material would be required.  However, there are some potential mechanisms for D/H enhancements at high temperature that might be active in these zones \citep{Thi2010MNRAS.407..232T} and future work will explore the HDO/H$_2$O (OD/OH) ratio of phyllosilicate grains.

\section{Conclusions \& perspective}\label{conclusions}

We explored the kinetics of formation of \phyllos\  from anhydrous silicate grains exposed either to water vapour or in contact with a water ice mantle from 10 to 900~K using a warm gas-grain surface chemical model. The \phyllos\ formation model was used in a zero-dimensional model and in a more realistic \pd\ model. We showed that different reservoirs of solid water are present in \pd. Water can be found in the cores of silicate grains (\phyllos) within radius of 10~au. Further away from the star, the temperatures are lower, and water can be frozen onto dust to form a thick mantle. The amount of water in \phyllos s can reach 2\%--3\% of the total mass of refractory material. The timescale for the formation of \phyllos s is shorter than the typical lifetime of \pd s of 2-3 Myrs. It proceeds by the adsorption to the silicate core followed by diffusion into the silicate core. The amount of water trapped as \phyllos\ can theoretically reach the maximum water storage capacity of silicates within the disks lifetime. Water can occupy the chemisorption sites required to efficiently form \H2. Formation of \H2 via hydrogenated PAHs and PAH cations becomes important depending on the abundance of PAHs in the \pd. Search for the infrared signatures of \phyllos s in \pd s is warranted to test the
model, which predicts that a significant amount of the silicates should be hydrated in the warm region (100-700~K). Our modelling supports the scenario, in which water found on Earth may have been already trapped in the dust grains at the phase of planetesimal formation. The amount of \phyllos s formed by this method may not be efficient enough compared to the widespread amount found in Solar System objects and subsequent aqueous alterations may be still required. 

\begin{acknowledgement}
We would like to thank the anonymous referee for the comments that improved the manuscript. We thank  Dr. Jake Laas, Dr. Thomas M\"{u}ller, and Dr. Victor Ali-Lagoa. IK,WFT, CR, and PW acknowledge funding from the EU FP7- 2011 under Grant
Agreement nr. 284405. CR also acknowledges funding by the Austrian Science
Fund (FWF), project number P24790. This research has made use of NASA?s Astrophysics Data System.
 \end{acknowledgement}

\bibliographystyle{aa} 
\bibliography{surface_chemistry,bringing_water_paper,hdo_h2o}

\begin{appendix}

\section{Hydrogen surface reactions}\label{H_surf_reactions}

Table~\ref{tab_dust_reactions} lists the grain surface and PAH reactions concerning atomic hydrogen.

\begin{table*}[!htbp]
\begin{center}
\caption{Main grain reactions involved in the formation and destruction of \H2. The energies are expressed in units of Kelvin.  \label{tab_dust_reactions}}
\vspace*{-3mm}          
\begin{tabular}{p{0.3cm}p{1cm}p{0.3cm}p{1cm}p{0.3cm}p{1cm}p{0.3cm}p{1cm}p{8cm}}     
\hline
\hline
\noalign{\smallskip}        
\multicolumn{8}{c}{Reaction} & \multicolumn{1}{c}{Comment}\\
\hline
\noalign{\smallskip}
1 &  & & H   & \rra & H\#    & & &     physisorption, barrierless\\
2 & H  & + & *    & \rra & *H\#   & &         & $E\gcH$~=~$\Eactc$~=~900~K \\
3 & H\#  &+&  *    & \rra & *H\#   & &        &  $E\pcH$~=~$\Eactc$ \\
\noalign{\smallskip}
\hline
\noalign{\smallskip}
4 & & &H\#        & \rra & H    & &       & \EbHp ~=~600~K \\
5 & H\# & + &     \UV     & \rra & H    & &       & photodesorption \\
6 & H\# & + &CR    & \rra & H    & &       &  \\
7 & & &*H\#        & \rra &  H   &+& *  & \EbHc~=~10,000~K \\
8 & *H\# & + & \UV     & \rra & H    & + &  *     &  \\
9 & *H\# & + &CR    & \rra & H    & + & *      &  via CR induced UV\\
\noalign{\smallskip}
\hline
\noalign{\smallskip}
10 & H   &+ &H\#   & \rra & \H2     & &        & Eley-Rideal (ER) mechanism, barrierless \\
11& H\#  &+& H\#   & \rra & \H2    & &        & $E\actHpHp$~=~0~K  \citep{ Navarro-Ruiz2014_C4CP00819G} \\
\noalign{\smallskip}
\hline
\noalign{\smallskip}
12 & H   &+ &*H\#  & \rra & \H2      & + &  *     &   ER mechanism, barrierless \\
13 & H\#   &+ &*H\#  & \rra & \H2      & + & *      &    $E\actHpHc$~=~$\Eactc$\\
14 & *H\# &+&  *H\# & \rra & \H2    & +& 2* & $E\actHcHc$~=~$2 \times \Eactc$\\
\noalign{\smallskip}
\hline
\noalign{\smallskip}
15 & \H2 & + & * & \rra & *H\# & + & H\# & \EdHHchem~=~3481~K \citep{Dino2004_713}\\ 
\noalign{\smallskip}
\hline
\noalign{\smallskip}
16 & & &  PAH        & \rra &  PAH\#    & &  &    physisorption \\
17 & & & HPAH        & \rra & HPAH\#    & &  &    -- \\
\noalign{\smallskip}
\hline
\noalign{\smallskip}
18 & & &  PAH\#         & \rra &  PAH    & &  &  desorption \\
19 & & & HPAH\#         & \rra & HPAH    & &  &     -- \\
20 & PAH\# &+ & \UV           & \rra &  PAH    & &  &  photodesorption  \\
21 & HPAH\#  &+ & \UV         & \rra & HPAH    & &  &    -- \\
22 & PAH\# &+ & CR          & \rra &  PAH    & &  &    cosmic-ray induced photodesorption\\
23 & HPAH\#  &+ & CR        & \rra & HPAH    & &  &    -- \\
\noalign{\smallskip}
\hline
\noalign{\smallskip}
24 &  H & + & PAH          & \rra & HPAH  & & &  $E\actPAHH$~=~692~K \citep{Rauls2008ApJ...679..531R} \\
25 & & & HPAH   & \rra & H & + & PAH  &  thermal H-detachement with \EbHPAH~=~16,250~K \\
26 &HPAH  & + & \UV    & \rra & H & + & PAH  & photodetachment, $E$(C-H)~=~4.45 eV (51640~K)  \\
27 & HPAH & + & H & \rra & \H2   & + & PAH  & cross-section $\sigma=$0.06 $\AA^2$/C atom, $E^{\rm act}$ = 0~K$^a$  \citep{Mennella2012ApJ...745L...2M}  \\
28 &  H  & + & PAH$^+$  & \rra & HPAH$^+$  & & &  $E\actPAHpH$~=~116~K\\
29 & HPAH$^+$ & + & e & \rra & PAH   & + & H  & dissociative recombination  \\
30 & HPAH$^+$ & + & H & \rra & PAH$^+$  & + &  \H2 & Langevin rate \citep{Montillaud2013AA...552A..15M}  \\
\noalign{\smallskip}
\hline
\noalign{\smallskip}
31 &  \H2  & + & PAH  & \rra & HPAH  & + & H & \EdHHPAH~=~3481~K\\
\noalign{\smallskip}
\hline
\noalign{\smallskip}
32 & PAH & + & \UV   & \rra & PAH$^+$    & + &   e  & photoionisation\\
33 & PAH     & + & e  & \rra & PAH$^-$    & &     & electron attachment\\
34 & PAH$^-$ & + & \UV   & \rra & PAH    &  + &   e  & photodetachement\\
35 & PAH$^+$ & + & e  & \rra & PAH    & &     & electron recombination \\
36 & PAH$^+$ & + & X  & \rra & PAH    & +  &  X$^+$   & charge exchange with species X \\
37 & PAH        & + & X$^+$  & \rra & PAH$^+$    & +  &  X & -- \\
\noalign{\smallskip}
\hline
\noalign{\smallskip}
38 &  H        & + & e$^{-}$ & \rra & H$^-$ & \UV  &  &radiative attachment\\
39 &  H        & + & H$^-$ & \rra & \H2    & + & e$^-$  &    associative detachment    \\ 
40 &  H + H  &+& H        & \rra & \H2    &+ & H       & three-body reactions\\
41 &  H + H   &+& \H2   & \rra & \H2    &+ & \H2     & -- \\
42 & \H2   & + & \UV   & \rra & H    & + & H  & photodissociation including self-shielding   \\
43 & \H2 & + & CR& \rra & H & + & H & by secondary electrons and CR-generated UV photons\\ 
\noalign{\smallskip}
\bottomrule
\end{tabular}
\resizebox{170mm}{!}{
\begin{minipage}{170mm}{
$E\gcH$ is the energy barrier for chemisorption of a gas-phase hydrogen atom on the grain surface. $E\pcH$ is the energy barrier for physisorbed H-atom to chemisorb on the grain surface.
\EbHp is the adsorption energy of a physisorbed H-atom. We assumed no activation barrier for 
\H2 recombination between two physisorbed H-atoms ($E\actHpHp$=0~K). 
$E\actHpHc$ and $E\actHcHc$ are activation energy for \H2 recombination involving one or two chemisorbed H-atom(s). \EdHHchem is the \H2 dissociative adsorption.}
\end{minipage}}
\end{center}
\end{table*}

\section{Disk model extra results}\label{disks_extra_results}

\begin{figure*}[!htbp]   
  \centering
  \includegraphics[angle=0,width=8.0cm,height=8cm,trim=50 80  80 300, clip]{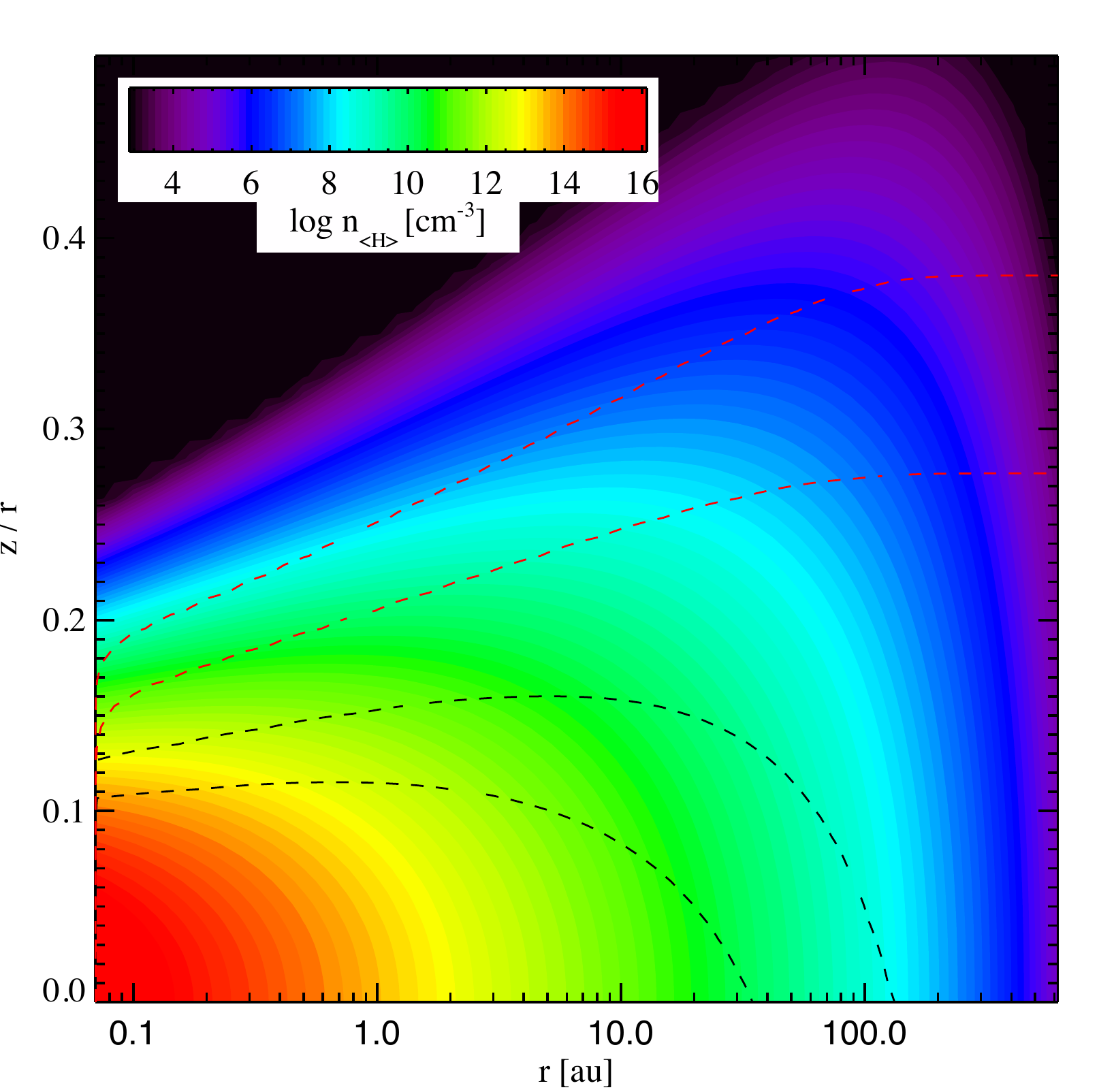}  
  \includegraphics[angle=0,width=8.0cm,height=8cm,trim=50 80  80 300, clip]{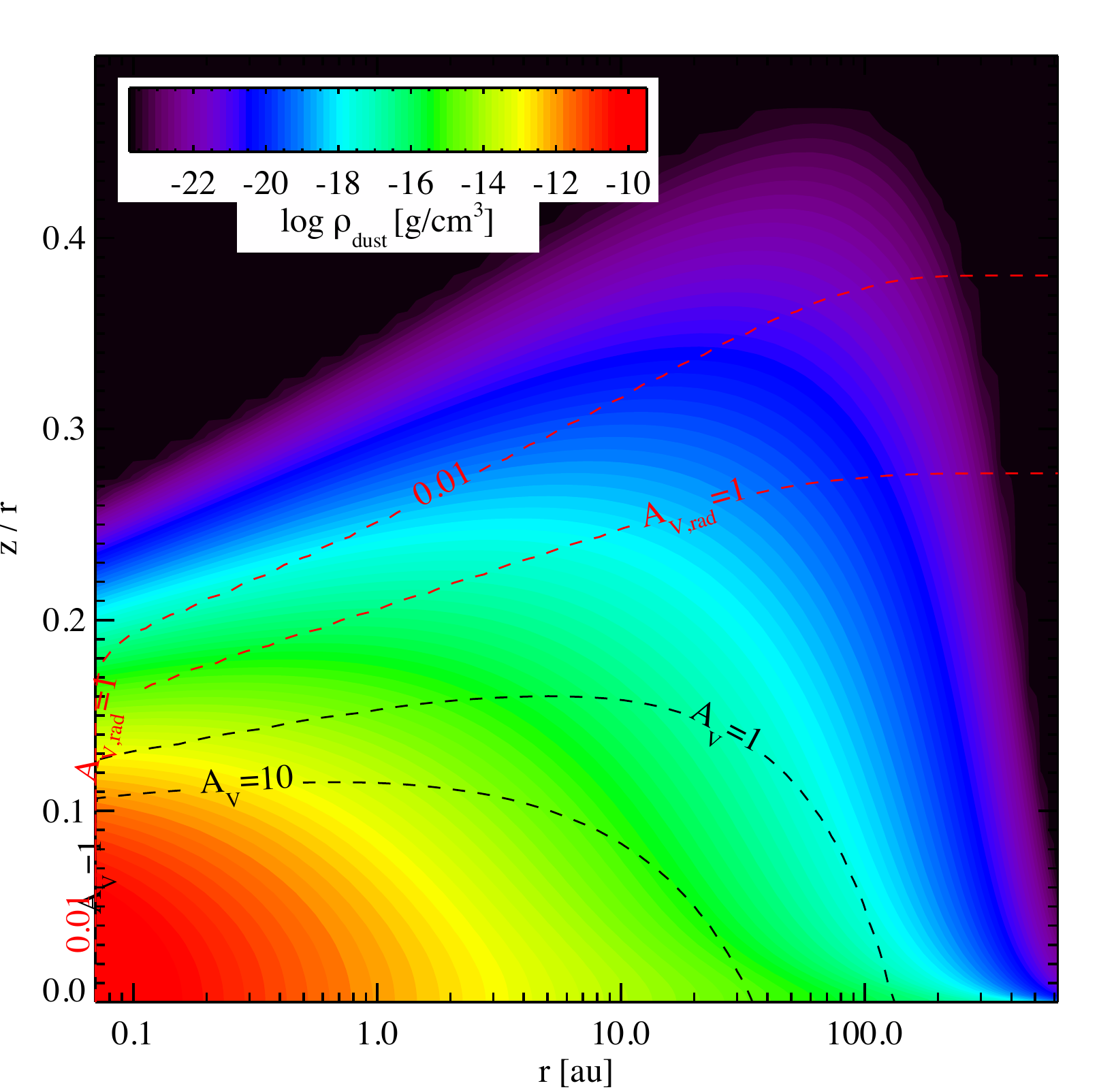}
  \includegraphics[angle=0,width=8.0cm,height=8cm,trim=50 80  80 300, clip]{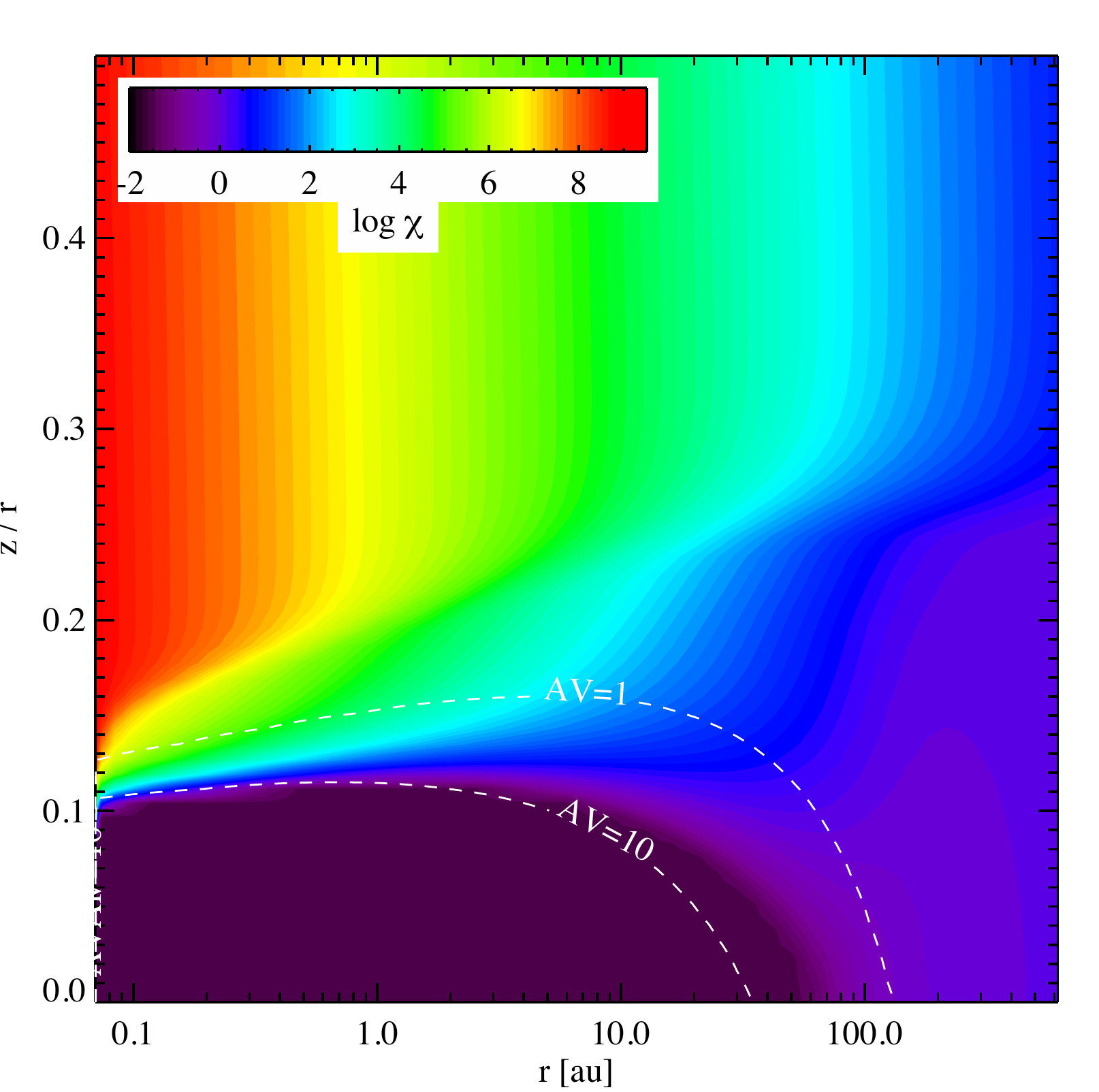}
  \includegraphics[angle=0,width=8.0cm,height=8cm,trim=50 80  80 300, clip]{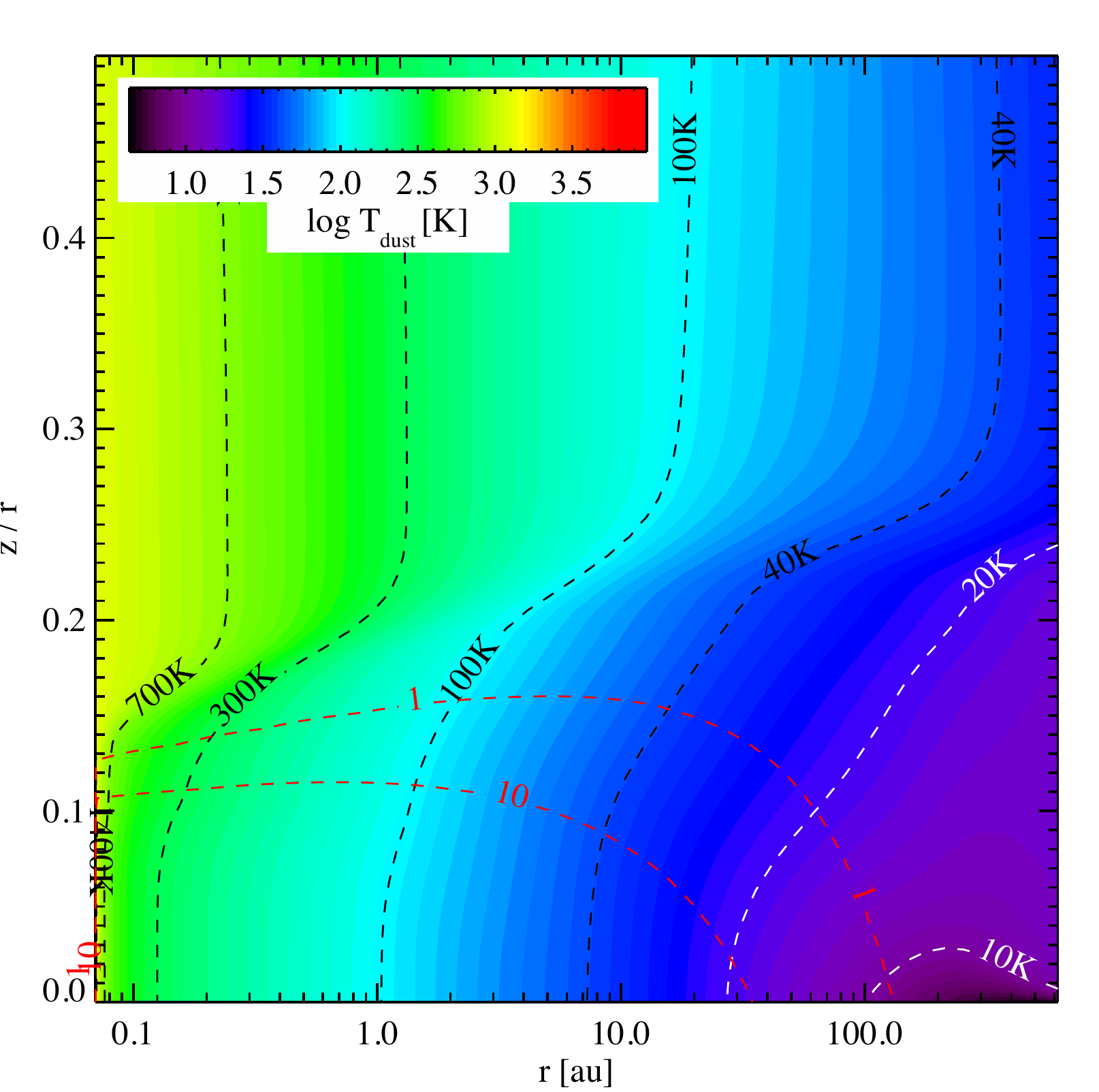} 
  \includegraphics[angle=0,width=8.0cm,height=8cm,trim=50 80  80 300, clip]{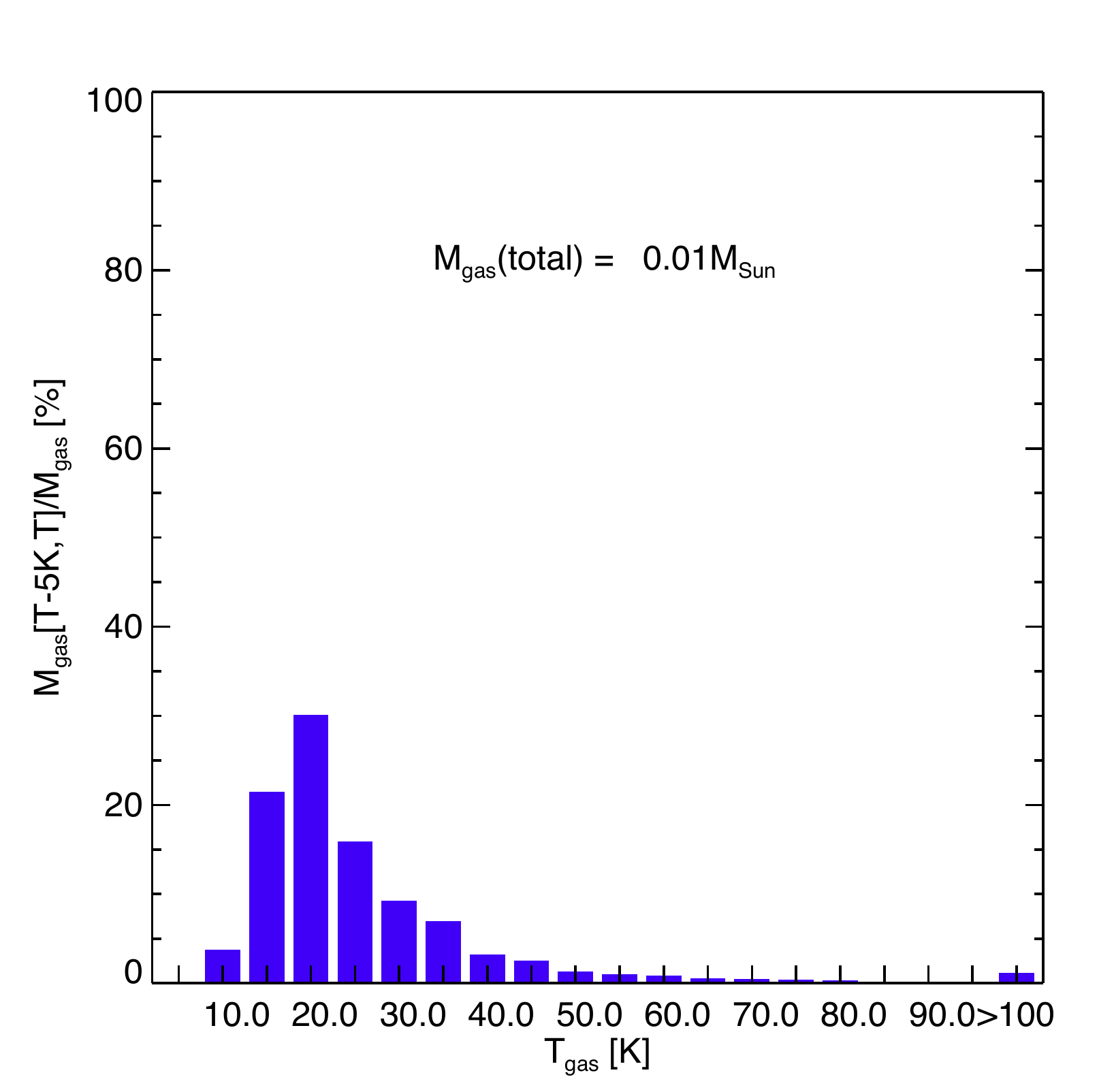}
  \includegraphics[angle=0,width=8.0cm,height=8cm,trim=50 80  80 300, clip]{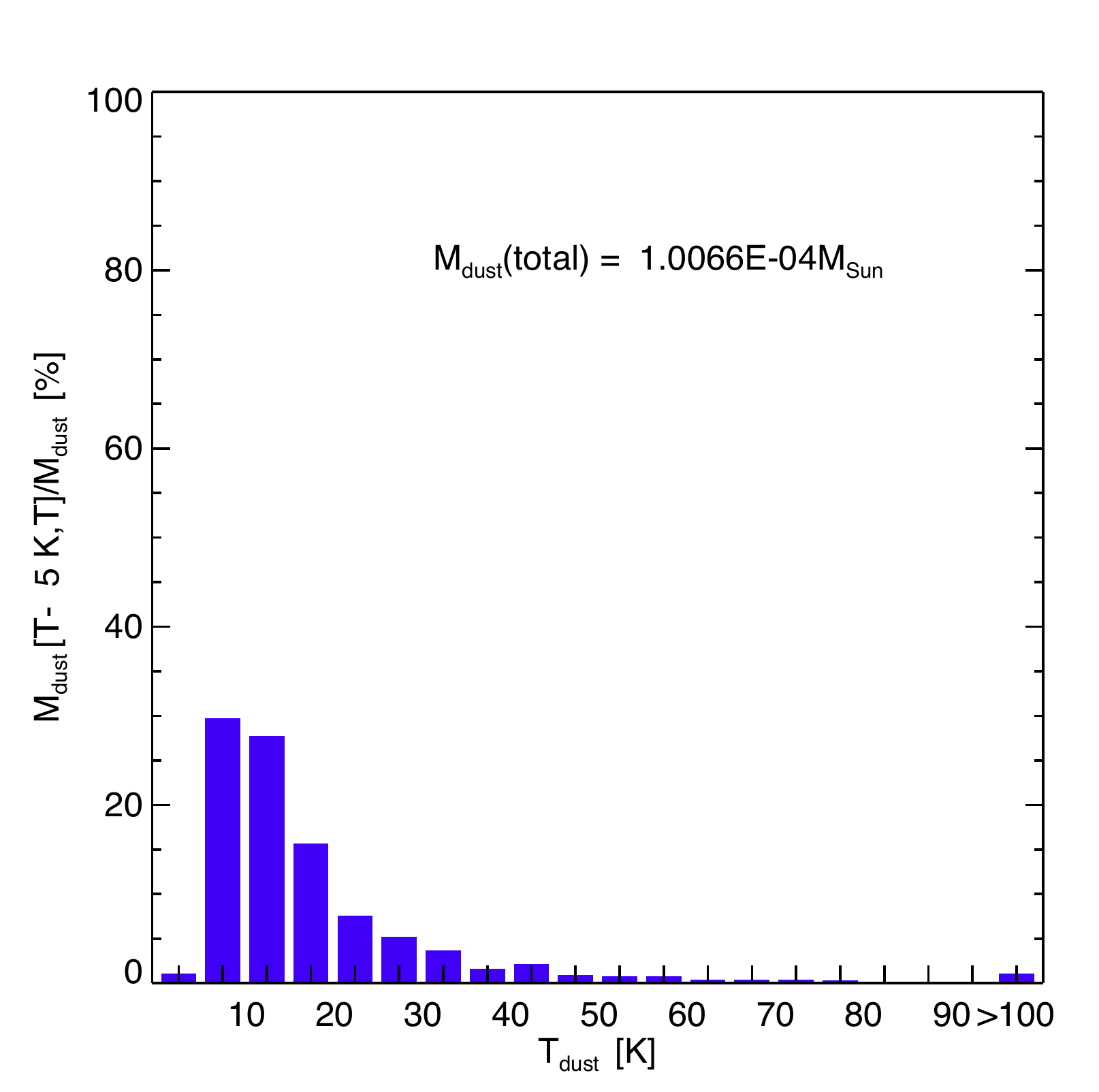} 
  \caption{The top two rows show the disk distribution of various quantities: disk gas number density ($n_{\rm H}$), dust mass density $\rho_{\rm dust}$, UV field units of $\chi$ ($\chi$~=~1 is the average in the local interstellar medium radiation field) over-plotted by contours showing the dust extinction in the optical, and the dust temperature. The lower panels show the gas and dust temperature distribution in the disk.}
  \label{fig_disk_results2}          
\end{figure*}

Fig.\ref{fig_disk_results2} show extra results from the disk model.

\section{Zero dimensional model at 10 K}\label{zeroD_10K_model}

\begin{figure*}[!ht]
  \centering  
   \includegraphics[angle=0,width=9.0cm,height=7.5cm,trim=25 70  70 300, clip]{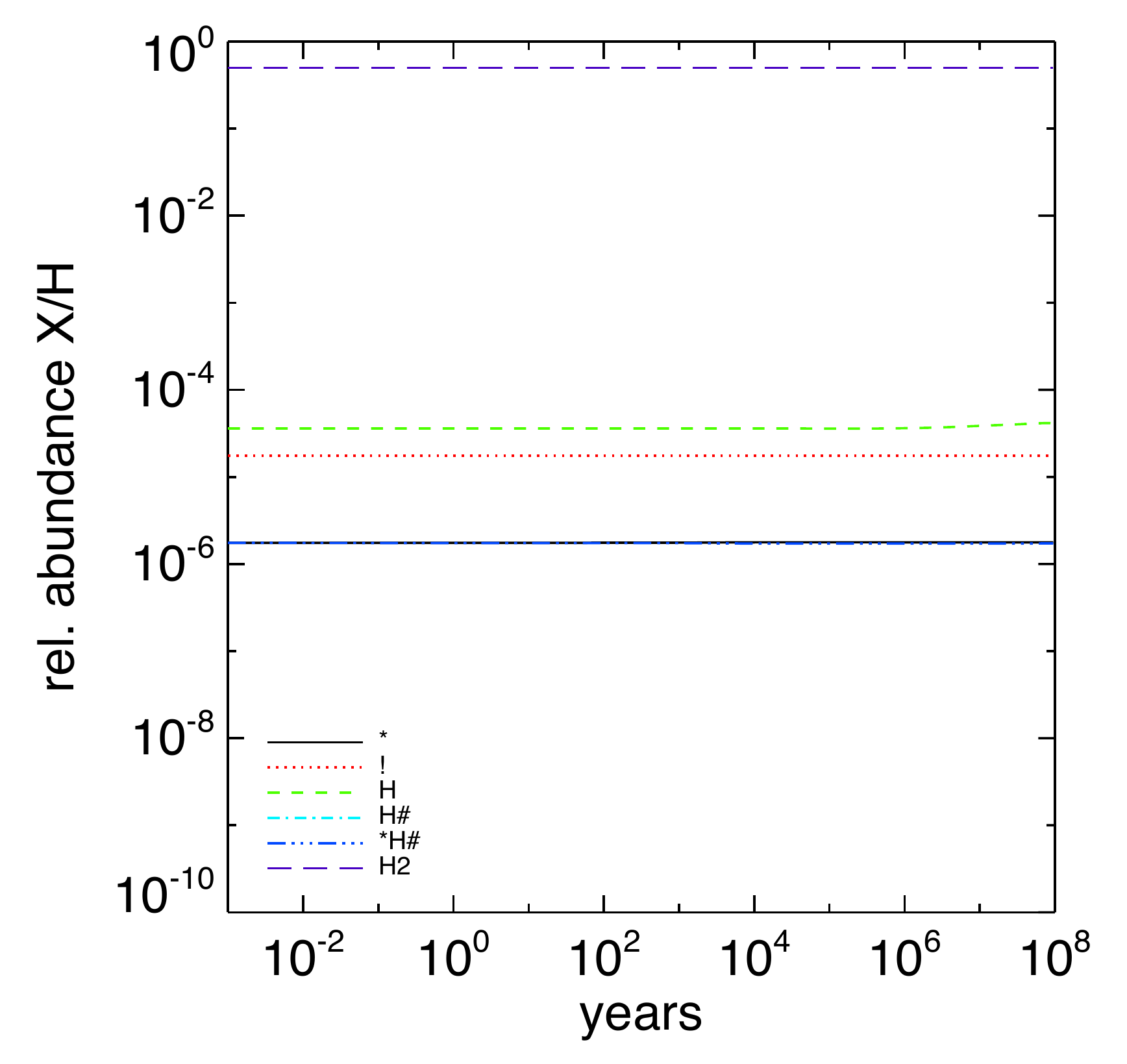}  
  \includegraphics[angle=0,width=9.0cm,height=7.5cm,trim=25 70  70 300, clip]{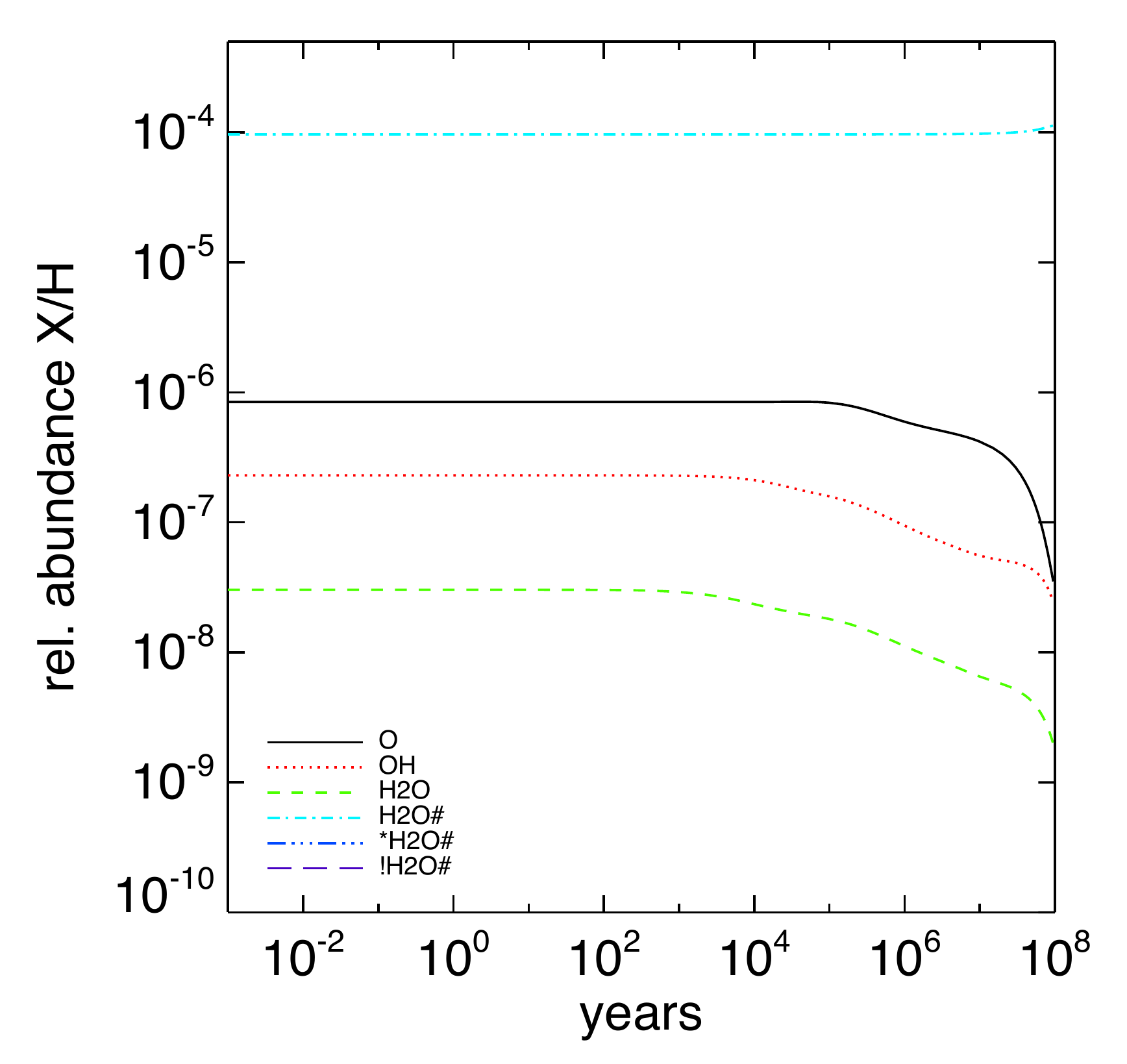}
    \caption{Species abundances as function of time at 10~K.  The abundance of unoccupied silicate surface (*) and core (!) sites, of atomic hydrogen in the gas and solid phases, and of \H2 are shown in the left panels. Gas-phase water and water ice abundances are shown on the right panels together with gas-phase abundance of atomic oxygen and OH.}
  \label{fig_water_abundances_time_10K}          
\end{figure*}

Fig.~\ref{fig_water_abundances_time_10K}  show results from the zero dimensional model at 10~K.

\section{Lizardite abundance from chemical equilibrium calculation}\label{lizardite}

\begin{figure}
  \centering  
   \includegraphics[angle=0,width=9.0cm,height=7.5cm,trim=45 70  80 300, clip]{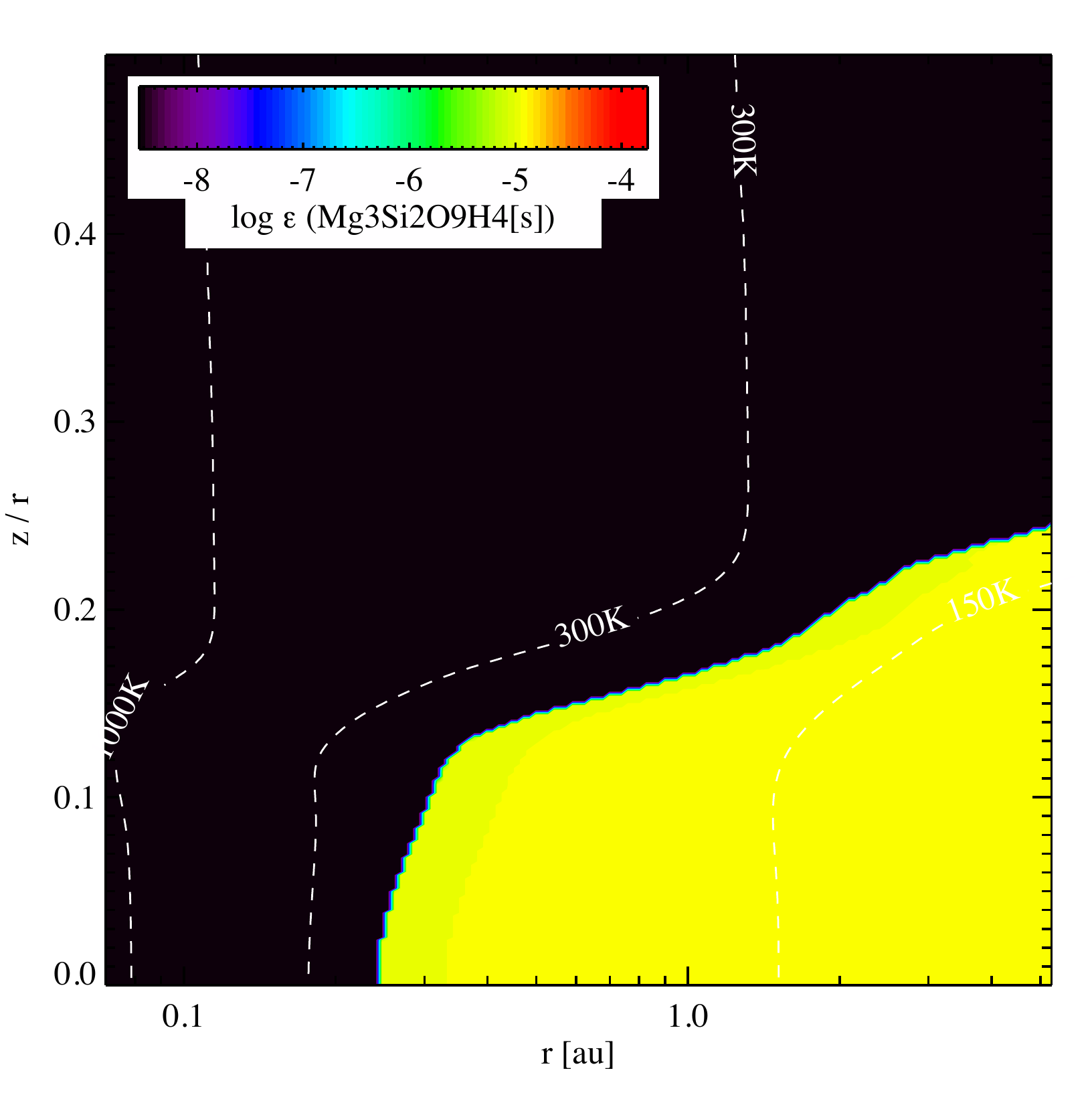}\label{fig_Lizardite_GGchem}   
  \caption{Lizardite Mg$_3$Si$_2$O$_9 $H$_4$ abundance in the inner disk as computed with the chemical equilibrium solver {\sc GGChem} \citep{Woitke2018A&A...614A...1W}.}
  \label{fig_Lizardite_GGchem}          
\end{figure}  

As an illustration that phyllosillicate is the most thermodynamically stable species in the inner disk midplane, we show the abundance of Lizardite Mg$_3$Si$_2$O$_9$H$_4$, the hydrated form of Olivine in Fig.~\ref{fig_Lizardite_GGchem}. The abundance has been computed with the chemical equilibrium code {\sc GGChem}, which is called by {\sc ProDiMo} when the temperature is above 100~K \citep{Woitke2018A&A...614A...1W}.

\end{appendix}

\end{document}